\documentclass[11pt]{article}

\usepackage{graphicx,amsmath,amsthm,mathrsfs,amssymb}
\usepackage[usenames]{color}
\usepackage{ulem}
\usepackage{authblk}
\usepackage{cite}
\usepackage{float}

\def\ltwid{\mathrel{\raise.3ex\hbox{$<$\kern-.75em\lower1ex\hbox{$\sim$}}}}
\def\gtwid{\mathrel{\raise.3ex\hbox{$>$\kern-.75em\lower1ex\hbox{$\sim$}}}}

\def\overleftrightarrow#1{\vbox{\ialign{##\crcr
     $\leftrightarrow$\crcr\noalign{\kern-1pt\nointerlineskip}
     $\hfil\displaystyle{#1}\hfil$\crcr}}}

\newcommand{\be}{\begin{equation}}
\newcommand{\ee}{\end{equation}}
\newcommand{\bea}{\begin{eqnarray}}
\newcommand{\eea}{\end{eqnarray}}
\newcommand{\nn}{\nonumber}

\newcommand\ovX{\overline{X}}
\newcommand\ovf{\overline{f}}
\newcommand\ovU{\overline{U}}

\begin{document}

\begin{titlepage}

\begin{flushright}
UFIFT-QG-18-06
\end{flushright}

\vskip 2cm

\begin{center}
{\bf Do the Spirits Rise?}
\end{center}

\vskip 1cm

\begin{center}
Sohyun Park$^{1\star}$ and R. P. Woodard$^{2\dagger}$
\end{center}

\vskip .5cm

\begin{center}
\it{$^{1}$ CEICO, Institute of Physics of the Czech Academy of Sciences, \\ 
Na Slovance 2, 18221 Prague 8 CZECH REPUBLIC}
\end{center}

\begin{center}
\it{$^{2}$ Department of Physics, University of Florida,\\
Gainesville, FL 32611, UNITED STATES}
\end{center}

\vspace{1cm}

\begin{center}
ABSTRACT
\end{center}
A nonlocal gravity model based on $\frac1{\square} R$ achieves the 
phenomenological goals of generating cosmic acceleration without dark 
energy and of suppressing the growth of perturbations compared to the 
$\Lambda$CDM model. Although the localized version of this model possesses
a scalar ghost, the nonlocal version does not suffer from any obvious 
problem with ghosts. Here we study the possibility that the scalar ghost
mode might be uncontrollably excited through time evolution, even though 
it is initially absent. We present strong evidence that this does not 
happen, so the analogy is to the conformal mode of general relativity
which can be excited but only in a controlled way.

\begin{flushleft}
\end{flushleft}

\vskip .5cm

\begin{flushleft}
$^{\star}$ e-mail: park@fzu.cz \\
$^{\dagger}$ e-mail: woodard@phys.ufl.edu
\end{flushleft}

\end{titlepage}

\section{Introduction}

The evidence is strong that our universe is currently undergoing a phase of
accelerated expansion \cite{SNIA-Riess, SNIA-Perlmutter, newSN, data1, data2, data3, data4, data5}, 
however, there is no similarly strong indication concerning the cause. 
The simplest solution is the $\Lambda$CDM model, in which acceleration is driven 
by a very small cosmological constant.
This model fits the observed expansion history when the fractional energy 
densities of the cosmological constant, of nonrelativistic matter and of
radiation take the respective values 
$
\Omega_\Lambda \approx 0.7 \;, \Omega_m \approx 0.3 \;, \Omega_r \approx 
8.5 \times 10^{-5} .
$
However, the $\Lambda$CDM model raises some theoretical concerns:
\begin{enumerate}
\item{Why is the energy density of the cosmological constant 
$\rho^{\rm obs}_{\Lambda} \sim (10^{-3} \mbox{eV})^4$ 
so small compared to the natural energy densities of fundamental theory?}
\item{Why does $\rho^{\rm obs}_{\Lambda}$ have a value which causes it to
become dominant so recently in cosmic history?}
\end{enumerate}
These are, respectively, the old and new problems of the cosmological constant 
\cite{CC-review-Weinberg, CC-review-Martin, CC-review-Burgess, CC-review-Padilla}.

There have also been extensive efforts to explain cosmic acceleration by 
modified gravity \cite{MG-review-CFPS, MG-review-CL, MG-review-NOO}. The only local, metric-based, generally 
coordinate invariant and potentially stable class of models are based on 
generalizing $R$ in the Einstein-Hilbert Lagrangian to $f(R)$ \cite{W-lecture}. 
However, these models can only reproduce the $\Lambda$CDM expansion history 
for the $\Lambda$CDM choice of $f(R) = R - 2\Lambda$ \cite{f(R)=R-2L}. Among 
the three remaining options of using fields other than the metric to carry 
part of the gravitational force, breaking general covariance or abandoning 
locality \cite{W-shenzhen,W-review-2014}, we consider a metric-based, invariant,
nonlocal modification based on distorting the Einstein-Hilbert Lagrangian by 
an algebraic function of the nonlocal scalar $\square^{-1} R$ defined with 
retarded boundary conditions \cite{DW-2007},
\be
S_{\rm DW} = \frac{1}{16\pi G}\int d^4 x \sqrt{-g} \left[ R + R 
f\Bigl(\frac{1}{\square} R\Bigr) \right]  \;.
\label{DW-action}
\ee
Because the nonlocal scalar $\square^{-1} R$ is dimensionless this class of 
models avoids the introduction of the new mass scale which is so problematic 
for the $\Lambda$CDM model. It also incorporates two features which naturally
delay the onset of cosmic acceleration to very late times:
\begin{itemize}
\item{Nothing happens during radiation domination because $R = 0$; and} 
\item{Even after matter domination the growth of $\square^{-1} R$ is only
logarithmic in the co-moving time, so that its current value is about $-14$.}
\end{itemize}
The algebraic function $f(X)$ can be chosen for negative $X$ to reproduce the 
$\Lambda$CDM expansion history \cite{DW-2009,EPV-1209,EPVZ-1302}. By taking 
$f(X)$ to vanish for positive $X$ one completely avoids the problems inside 
gravitationally bound systems that are so challenging for $f(R)$ models.

Closely related nonlocal models have also exploited the delayed response of 
$\square^{-1} R$ \cite{VAAS-2017,MM-2014,ABN-1707}:
\bea
S_{\rm MM} &=& \frac{1}{16\pi G}\int d^4 x \sqrt{-g} \left[ R - m^2 R 
\frac{1}{\square^2} R \right] \;, \label{LMM} \\
S_{\rm VAAS} &=& \frac{1}{16\pi G}\int d^4 x \sqrt{-g} \left[ R + m^2
\frac{1}{\square} R \right] \;, \label{VAAS} \\
S_{\rm ABN} &=& \frac{1}{16\pi G}\int d^4 x \sqrt{-g} \left[ R - m^4 
\frac{1}{\square^2} R \right] \;. \label{ABN}
\eea
These models approximately reproduce the $\Lambda$CDM expansion history,
however, they all require a new mass parameter $m^2$ which of the same 
order as the $\Lambda$CDM cosmological constant. In each of the models 
(\ref{DW-action}-\ref{ABN}) perturbations about the cosmological background
show deviations from the $\Lambda$CDM model. For (\ref{DW-action}) the
growth rate on the largest scales is reduced, relative to the $\Lambda$CDM
model, which improves the fit to existing data \cite{NCA-2017,spark-2017}. 
The trend is opposite for (\ref{LMM}), although not enough to falsify the
model \cite{Maggiore-1403,BLHBP-1408,Maggiore-1411,Maggiore-1602,Dirian-1704,BDFM-1712}.

Each of the nonlocal models (\ref{DW-action}-\ref{ABN}) can be re-cast in
a localized form by the introduction of auxiliary scalar fields. For the
original model (\ref{DW-action}) the localized version employs scalar
fields $X$ and $U$ \cite{NO-2007, CMN-2008, Koivisto-0803, Koivisto-0807, 
Koshelev-2008, NOSZ-1010, ZS-1108, FeliceSasaki-1412},
\be
S= \frac{1}{16\pi G}\int d^4 x \sqrt{-g} \Bigl[ R + R f(X) + g^{\mu\nu}
\partial_\mu X\partial_\nu U + UR\Bigr]  \;.
\label{DW-action-localized-IBP}
\ee
Varying with respect to $U$ and $X$ and substituting the solutions in
(\ref{DW-action-localized-IBP}) seems to recover the original, nonlocal
form (\ref{DW-action}),
\bea
\frac{16\pi G}{\sqrt{-g}}\frac{\delta S}{\delta U} & = & -\square X + R = 0 
\quad \Rightarrow \quad X = \frac{1}{\square} R + X_{\rm homo} \;, 
\label{Xeqn} \\
\frac{16\pi G}{\sqrt{-g}}\frac{\delta S}{\delta X} & = & Rf'(X) -\square U = 0 
\quad \Rightarrow \quad U = \frac{1}{\square} \Bigl(R f'(X)\Bigr) + U_{\rm homo} 
\;, \label{Ueqn}
\eea
where $\square X_{\rm homo} = 0 = \square U_{\rm homo}$. Whereas the localized
model (\ref{DW-action-localized-IBP}) could be regarded as a fundamental theory,
which might be subjected to quantization, the presence of the inverse d`Alembertian
in the original, nonlocal model (\ref{DW-action}) means that it can only be treated
as an effective field theory. In fact it was proposed to represent the most
cosmologically significant part of the quantum gravitational effective action
induced by graviton loops during primordial inflation \cite{DW-2007}.

Another important difference between the localized model and its nonlocal
ancestor concerns degrees of freedom. The localized version 
(\ref{DW-action-localized-IBP}) contains two scalar degrees of freedom, 
corresponding to the arbitrary initial value data which determine the 
homogeneous solutions $X_{\rm homo}$ and $U_{\rm homo}$ in relations
(\ref{Xeqn}) and (\ref{Ueqn}). In the original, nonlocal version 
(\ref{DW-action}) the fields $X$ and $U$ obey retarded boundary conditions, 
that is, both they and their first time derivatives vanish on the initial 
value surface. Hence the nonlocal model (\ref{DW-action}) lacks the two 
extra scalar degrees of freedom which are present in its localized 
counterpart (\ref{DW-action-localized-IBP}). This difference is crucial 
because the field redefinition $A_{\pm} = \frac12 (X \pm U)$ reveals that 
$A_{+}$ is a ghost field,
\be
g^{\mu\nu}\partial_\mu X\partial_\nu U =  g^{\mu\nu} \Bigl[\partial_\mu A_{+}
\partial_\nu A_{+}  - \partial_\mu A_{-}\partial_\nu A_{-}\Bigr]\;.
\label{kinetic-term-diagonalized}
\ee
Relation (\ref{kinetic-term-diagonalized}) has two important consequences
\cite{DW-2013,W-review-2014}:
\begin{itemize}
\item{The original, nonlocal model (\ref{DW-action}) is a 
constrained version of the localized model (\ref{DW-action-localized-IBP})
in which the scalars $X$ and $U$ and their first derivatives vanish on the
initial value surface; and}
\item{The localized model (\ref{DW-action-localized-IBP}) suffers from a
kinetic energy instability whereas the original, nonlocal model 
(\ref{DW-action}) may be stable.}
\end{itemize}

In a stable theory one can only excite one degree of freedom by lowering 
the excitation of some other degree of freedom. Because there is only a
finite amount of energy available in any given initial system, there is
an upper limit to the wave number of a mode which can be excited. In 
contrast, interacting field theories with a kinetic energy instability, 
such as (\ref{DW-action-localized-IBP}), are driven to a peculiar time 
evolution in which negative energy modes of arbitrarily high wave number 
are excited, along with corresponding positive energy degrees of freedom. 
The conformal mode of general relativity would engender precisely such a 
kinetic instability were it not constrained to be nondynamical. The 
original, nonlocal model (\ref{DW-action}) has a chance of avoiding the 
kinetic instability because the ghost mode of its local counterpart 
(\ref{DW-action-localized-IBP}) is similarly constrained to be nondynamical. 
The purpose of this paper is to check that it stays that way. That is, we
seek to confirm that the evolution of permitted perturbations does not 
lead to explosive excitation of the ghost mode.

Note that we are {\it not} claiming the ghost mode remains zero, any 
more than stability proofs of general relativity require the conformal
factor to remain unity. In fact the ghost mode is nonzero even in the 
background solution \cite{DW-2009,EPV-1209,EPVZ-1302}, just as the
conformal factor of general relativity expands in the cosmological
background.\footnote{Another parallel between the ghost mode and the 
conformal factor of general relativity is that they can both be fixed
by a gauge choice. Of course employing such a gauge in no way avoids 
the instability that would result without initial value constraints.}
We will show that perturbations of the ghost field also become nonzero
but that they do so in a controlled way.

If we had an energy functional the task would be simple: we would merely
establish that the Hamiltonian is bounded below. Unfortunately, there
is no energy functional for gravitating systems in cosmology. What we will
do instead is to follow the evolution of scalar plane wave perturbations 
about the cosmological background, both with $X$ and $U$ obeying retarded 
boundary conditions and with them obeying a variety of more general initial 
conditions. Of course retarded boundary conditions correspond to the 
original, nonlocal model (\ref{DW-action}), whereas more general initial 
conditions access the ghost mode. The radical contrast between these two 
cases provides strong evidence that no ghost appears in the original, 
nonlocal model (\ref{DW-action}).

This paper has four sections, of which this Introduction is the first. In
section 2 we give the linearized field equations for scalar plane wave
perturbations in cosmology. Section 3 presents the results of numerical 
evolution from various initial conditions. Our conclusions comprise section
4.

\section{Cosmological scalar perturbations}

The field equations of any metric-based modification to gravity can be 
expressed as, 
\be
G_{\mu\nu} + \Delta G_{\mu\nu} = 8\pi G T_{\mu\nu} \;,
\label{nonlocal_field_eq}
\ee
where $G_{\mu\nu}$ and $T_{\mu\nu}$ are the usual is Einstein tensor and
stress-energy tensor, respectively. The modification appropriate to 
(\ref{DW-action}) is,
\bea
\lefteqn{\Delta G_{\mu\nu} =
\Bigl[ G_{\mu\nu} + g_{\mu\nu}\square - D_{\mu}D_{\nu} \Bigr]
\biggl\{\! f\Big(\frac{1}{\square}R\Big)  + \frac{1}{\square}\Bigl[R f'
\Big(\frac{1}{\square}R\Big)\Bigr] \!\biggr\}
} \nn \\
&& \hspace{1.5cm} + 
\Bigl[ \delta_{\mu}^{(\rho}\delta_{\nu}^{\sigma)} 
- \frac{1}{2}g_{\mu\nu}g^{\rho\sigma}\Bigl] 
\partial_{\rho} \Big(\frac{1}{\square}R\Big)
\partial_{\sigma}
\biggl(\frac{1}{\square}\Bigl[R f'
\Big(\frac{1}{\square}R\Big)\Bigr]\biggr) \;,
\label{DeltaGmn}
\eea
where $\square^{-1}$ is always defined with retarded boundary conditions.
The analogous localized form is,
\be
\Delta G_{\mu\nu} =
\Bigl[G_{\mu\nu} + g_{\mu\nu}\square - D_{\mu}D_{\nu} \Bigr] 
\Bigl[f(X) + U\Bigr] + \Bigl[ \delta_{\mu}^{(\rho}\delta_{\nu}^{\sigma)} 
\!-\! \frac{1}{2}g_{\mu\nu}g^{\rho\sigma}\Bigl] \partial_{\rho} X 
\partial_{\sigma} U\; . \label{DeltaGmn-local}
\ee
Recall again that (\ref{DeltaGmn-local}) only agrees with (\ref{DeltaGmn}) 
when the scalars $X$ and $U$ and their first derivatives vanish on the 
initial value surface. We will use this version of the theory, first 
with retarded boundary conditions and then with more general initial
conditions.

We consider scalar metric perturbations (in Newtonian gauge) about a homogeneous, 
isotropic and spatially flat background,
\be
ds^2 = -\Bigl[1 + 2\widetilde{\Psi}(t,\vec{x}) \Bigr] dt^2 +a^2(t) \Bigl[1 + 
2 \widetilde{\Phi}(t,\vec{x}) \Bigr] d\vec{x}\cdot d\vec{x} \; . \label{FLRW-metric}
\ee
The corresponding auxiliary scalars take the form,
\begin{equation}
X(t,\vec{x}) = \ovX(t) + \widetilde{X}(t,\vec{x}) \qquad , \qquad U(t,\vec{x}) =
\ovU(t) + \widetilde{U}(t,\vec{X}) \; .
\end{equation}
Each of the tilde-carrying perturbation fields can be decomposed into spatial plane 
waves,
\begin{eqnarray}
\widetilde{\Psi}(t,\vec{x}) = \int \frac{d^3k}{(2\pi)^3} \, e^{i \vec{k} \cdot 
\vec{x}} \Psi(t,\vec{k}) \qquad & , & \qquad 
\widetilde{\Phi}(t,\vec{x}) = \int \frac{d^3k}{(2\pi)^3} \, e^{i \vec{k} \cdot 
\vec{x}} \Phi(t,\vec{k}) \; , \\
\widetilde{X}(t,\vec{x}) = \int \frac{d^3k}{(2\pi)^3} \, e^{i \vec{k} \cdot 
\vec{x}} \delta X(t,\vec{k}) \qquad & , & \qquad 
\widetilde{U}(t,\vec{x}) = \int \frac{d^3k}{(2\pi)^3} \, e^{i \vec{k} \cdot 
\vec{x}} \delta U(t,\vec{k}) \; .  
\end{eqnarray}
We use a slightly different notation for the energy density,
\begin{equation}
T_{00}(t,\vec{x}) = \rho(t) + \widetilde{\rho}(t,\vec{x}) = \rho(t) \Biggl[1 +
\int \frac{d^3k}{(2\pi)^3} \, e^{i \vec{k} \cdot \vec{x}} \delta(t,\vec{k}) 
\Biggr] \; .
\end{equation}
Because each spatial plane wave evolves independently at linearized order we will
simply give equations for the Fourier components $\Psi(t,\vec{k})$, $\Phi(t,\vec{k})$,
$\delta X(t,\vec{k})$, $\delta U(t,\vec{k})$ and $\delta(t,\vec{k})$. 

It remains to give the equations for the background quantities, and for the linearized 
perturbation fields. We first define the background nonlocal distortion function and
its first derivative,
\begin{equation}
\ovf \equiv f\Bigl(\ovX(t)\Bigr) \qquad , \qquad \ovf' \equiv f'\Bigl(\ovX(t) \Bigr) \; .
\end{equation}
The modified background metric field equations (\ref{nonlocal_field_eq}) are, 
\bea
3H^2 + [3H^2 + 3H\partial_{t}](\ovf +\ovU) +\frac{1}{2}\partial_t{\ovX}\partial_t{\ovU}
&=& 8\pi G \rho \;, 
\label{zeroth-00-local-eq}
\\
-(2\dot{H} + 3H^2) -[2\dot{H} + 3H^2 +2H\partial_{t}+\partial_{t}^2](\ovf +\ovU)
+\frac{1}{2}\partial_t{\ovX}\partial_t{\ovU}
&=& 8\pi G p \;,
\label{zeroth-trace-local-eq}
\eea
and the background auxiliary scalar field equations (\ref{Xeqn}-\ref{Ueqn}) are, 
\bea
-(\partial_t^2 + 3H\partial_t)\ovX&=& 6(\dot{H} +2H^2) \;, 
\label{eq:boxx-local}
\\
-(\partial_t^2 + 3H\partial_t)\ovU&=& 6(\dot{H} +2H^2) \ovf'
\;.
\label{eq:boxu-local}
\eea
The background fields $\ovX(t)$ and $\ovU(t)$, and their first
derivatives, vanish on the initial value surface.
In the sub-horizon regime of $k \gg H a$ the equations for linearized 
perturbations are \cite{spark-2017},
\bea
k^2\Phi +  k^2\left[\Phi(\ovf + \ovU) + \frac{1}{2}( 
\ovf' \delta X + \delta U)\right]  &\!=\!& 4\pi G a^2 \rho 
\delta \;,
\label{modified-Poisson-eq-summary}
\\
(\Phi+\Psi)+ (\ovf' \delta X + \delta U)+(\Phi+\Psi)(\ovf + \ovU) &\!=\!& 0\;,
\label{modified-gslip-eq-summary}
\\
\ddot{\delta} + 2H\dot{\delta} &\!=\!& -\frac{k^2}{a^2}\Psi \;,
\label{delta-evolution-eq-summary}
\\
\Bigl(-\partial_t^2 - 3H \partial_t -\frac{k^2}{a^2} \Bigr) \delta X 
& \! = \! & 2\frac{k^2}{a^2}(\Psi+2\Phi)\;,
\label{deltaX-local-sub-horizon-summary}
\\
\Bigl(-\partial_t^2 - 3H \partial_t -\frac{k^2}{a^2} \Bigr) \delta U 
& \! = \! & 2\frac{k^2}{a^2}(\Psi+2\Phi) \ovf' \;.
\label{deltaU-local-sub-horizon-summary}
\eea

\section{Perturbation growth with and without the ghost}

The purpose of this section is to compare the evolution of scalar 
perturbations in the original, nonlocal model (\ref{DW-action}) ---
which may be stable --- with perturbations in the localized theory 
(\ref{DW-action-localized-IBP}) --- which is certainly not stable. 
In both cases the evolution equations are 
(\ref{modified-Poisson-eq-summary}-\ref{deltaU-local-sub-horizon-summary});
the difference between the two models is the initial conditions 
obeyed by $\delta X(t,\vec{k})$ and $\delta U(t,\vec{k})$. The initial
conditions corresponding to the original, nonlocal model (\ref{DW-action}) 
are that these fields and their first derivatives vanish on the initial 
value surface. We first evolve from retarded boundary conditions, then
explore a variety of more general conditions, and finally contrast the
results.

The actual evolution is performed with respect to the cosmological
redshift,
\begin{equation}
1 \!+\! z \equiv \frac{a_{\rm now}}{a(t)} \Longrightarrow \frac{d}{dt} =
-(1 \!+\! z) H(z) \frac{d}{dz} \quad \mbox{where} \quad 
H(z) = H_0 \sqrt{\Omega_r(1+z)^4 + \Omega_m (1+z)^3 + \Omega_\Lambda} \;.
\end{equation}
The function $f(X)$ was chosen to make the expansion history exactly
that of the $\Lambda$CDM model \cite{DW-2009,EPV-1209,EPVZ-1302}.
We keep the initial conditions for the perturbations $\Phi(t,\vec{k})$, 
$\Psi(t,\vec{k})$ and $\delta(t,\vec{k})$ the same for all the cases:
\bea
\Phi(z_i) &\!\!\!=\!\!\!& \Phi_{\rm GR}(z_i)\;, ~
\Psi(z_i) = \Psi_{\rm GR}(z_i) = -\Phi_{\rm GR}(z_i)\;, \\ 
\delta(z_i) &\!\!\!=\!\!\!& \delta_{\rm GR}(z_i) = 
\frac{2k^2 a(z_i)}{3H_0^2 \Omega_{m}}\Phi_{\rm GR}(z_i)\;, ~ 
\delta'(z_i) = \delta'_{\rm GR}(z_i)\;.  
\eea
We set $z_i = 9$ (corresponding to $t_i = 0.55~{\rm Gyrs}$) and $k=100 H_0 
=  0.03h \rm{Mpc}^{-1}$ as in \cite{PD-2012,DP-2013,PS-2016,spark-2017}. 
By choosing $z_i = 9$ we can safely set the initial conditions for $\Phi$, 
$\Psi$ and $\delta$ the same as in general relativity (GR) because the 
nonlocal modifications are negligible for $z > 5$ \cite{DW-2009}. The 
scale of $k = 100H_0$ is small enough to take the subhorizon limit 
(or the quasi-static limit) and large enough to keep the perturbations 
linear \cite{PD-2012,DP-2013}.

\subsection{Perturbations without the ghost}

\begin{figure*}[htbp]
\begin{center}
 \begin{tabular}{cc} 
  \includegraphics[width=0.45\textwidth]{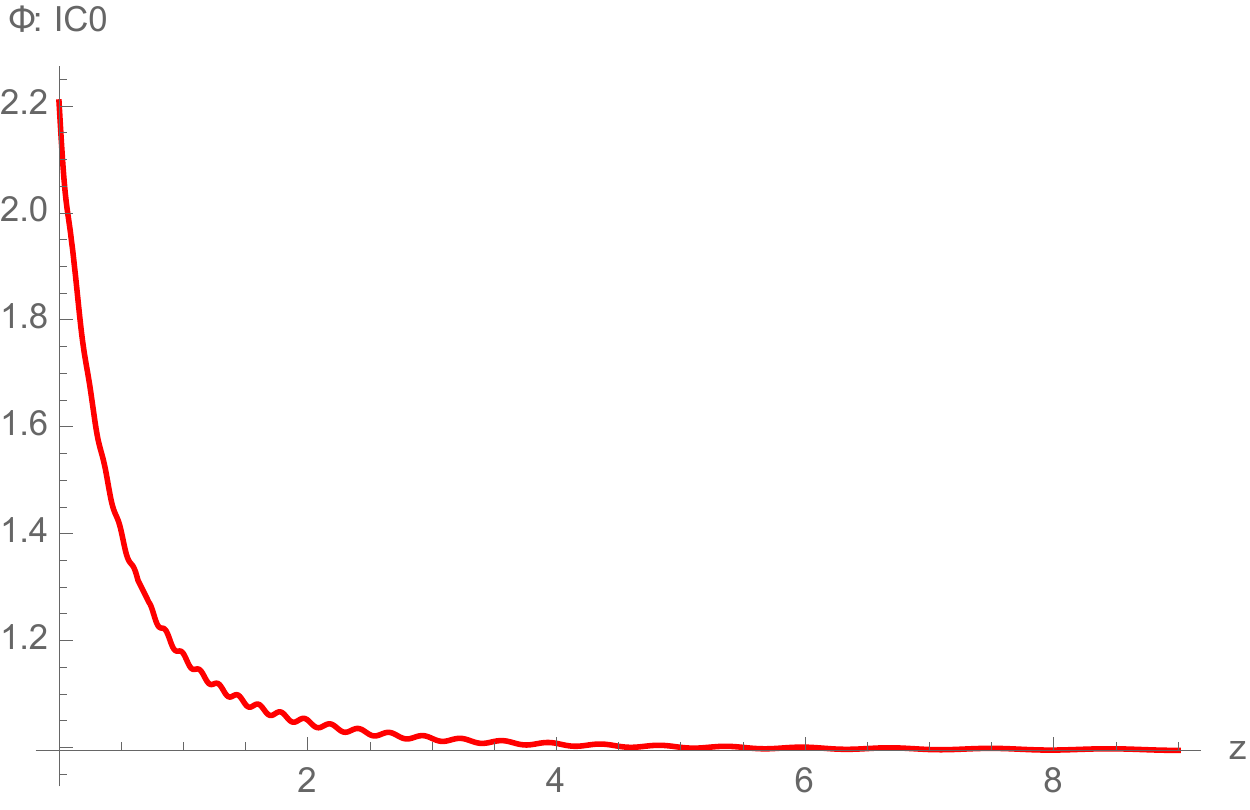}
  &
  \includegraphics[width=0.45\textwidth]{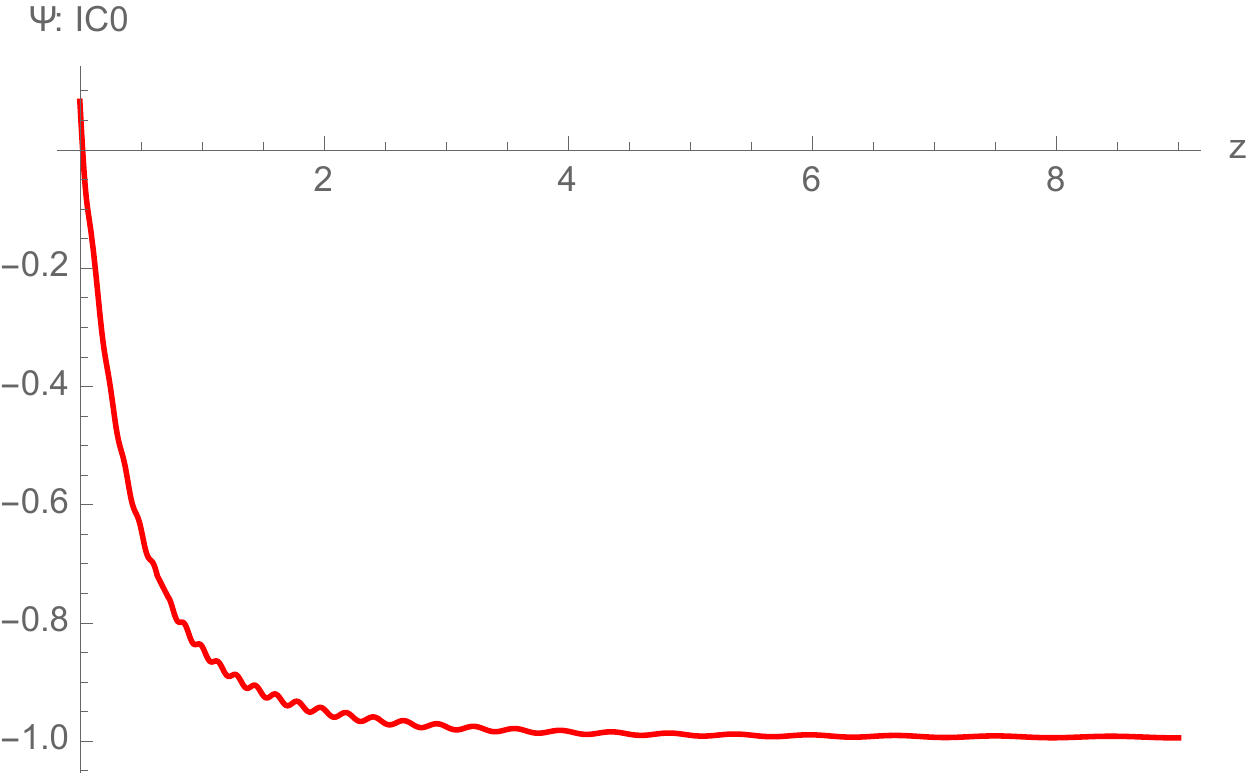}
\end{tabular}
\end{center}
\begin{center}
 \begin{tabular}{c} 
  \includegraphics[width=0.45\textwidth]{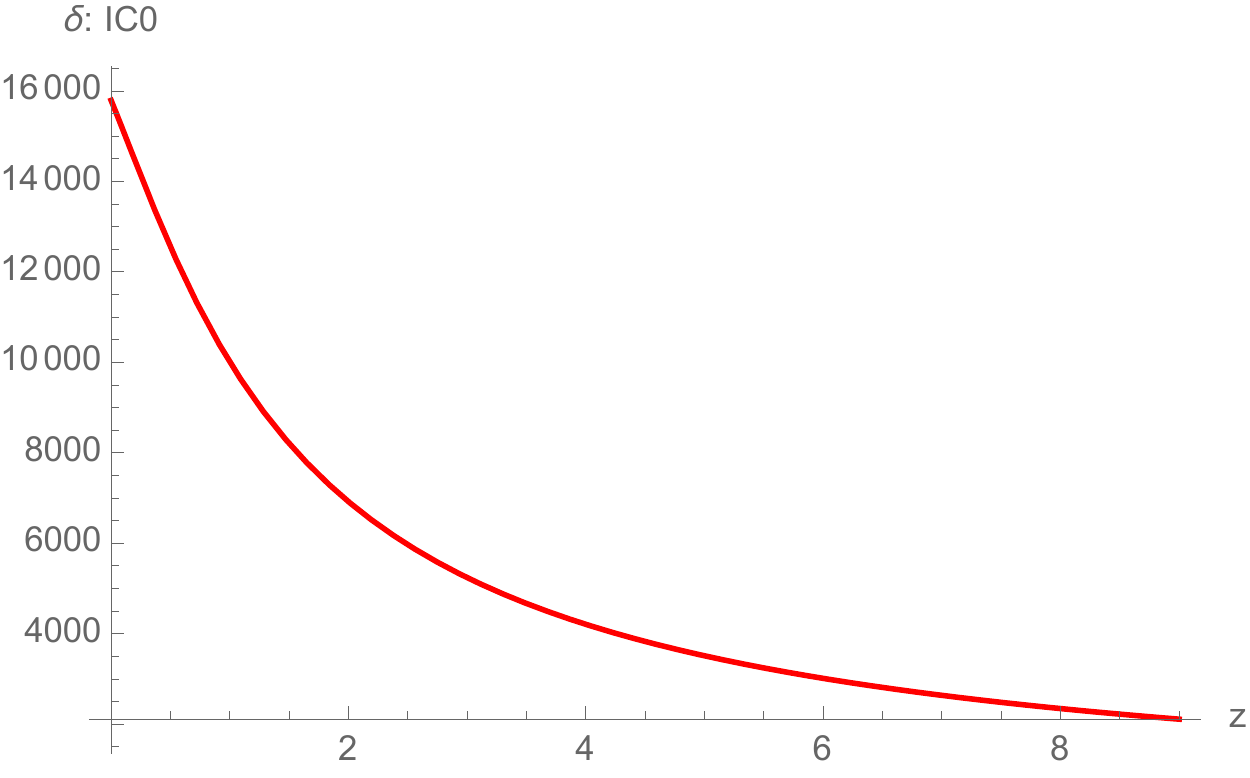} 
 \end{tabular}
\end{center}
\begin{center}
 \begin{tabular}{cc} 
  \includegraphics[width=0.45\textwidth]{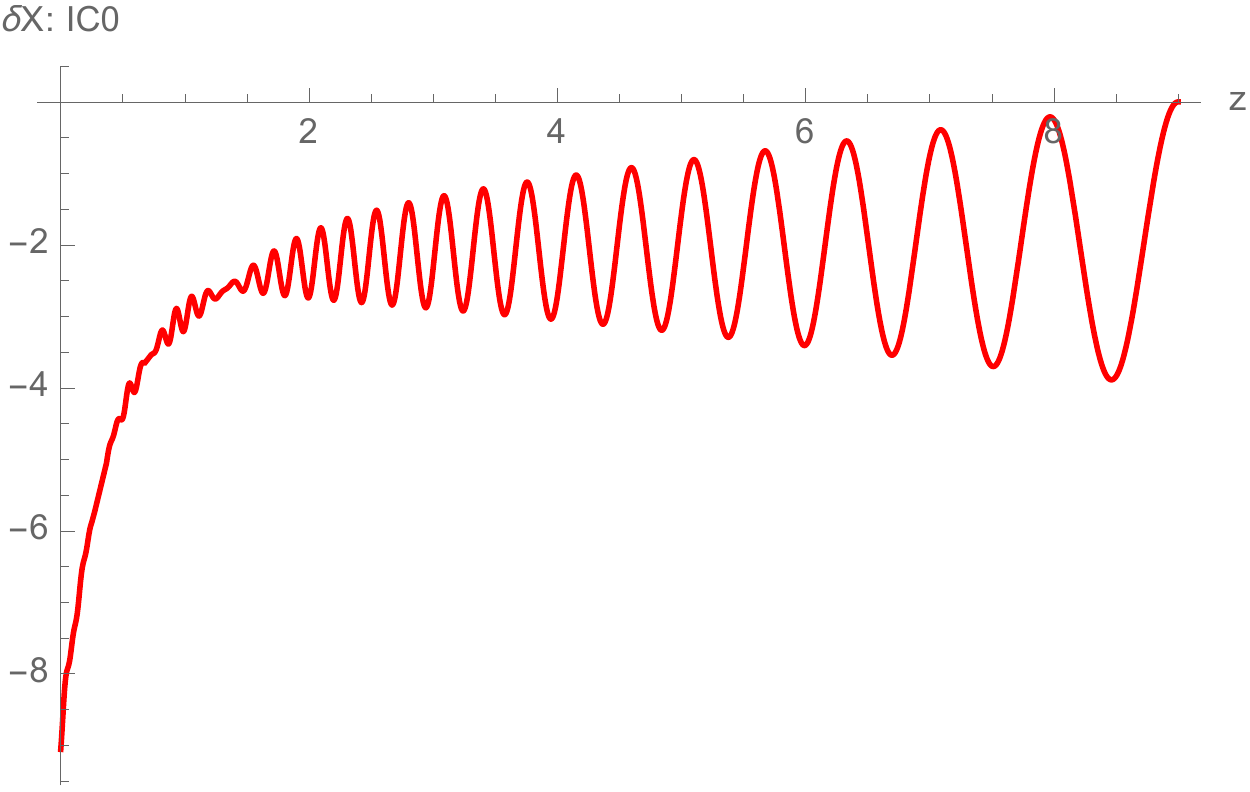}
  &
  \includegraphics[width=0.45\textwidth]{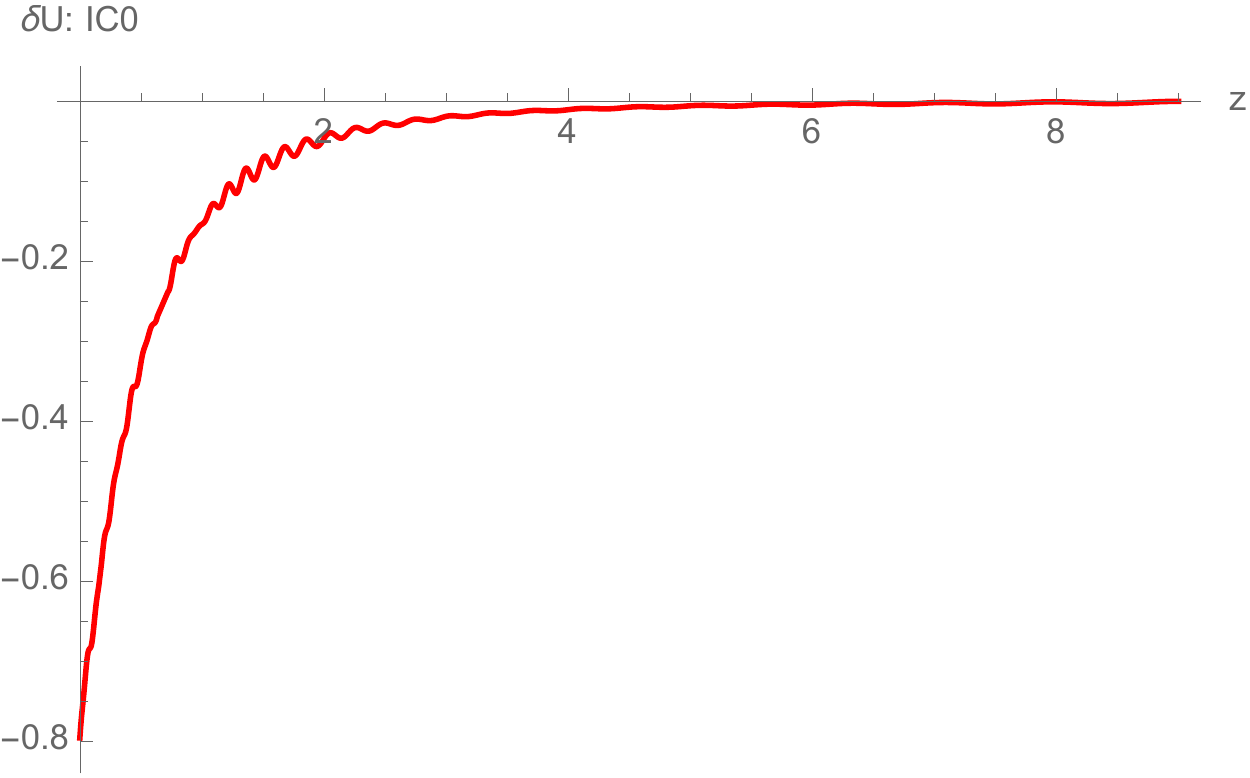}
 \end{tabular}
\end{center}
\caption{The evolution (as a function of redshift) of $\Phi(t,\vec{k})$, 
$\Psi(t,\vec{k})$, $\delta(t,\vec{k})$, $\delta X(t,\vec{k})$ and 
$\delta U(t,\vec{k})$ starting from IC0 initial conditions 
(\ref{nonlocal-IC}).} 
\label{fig:ic0-solutions}
\end{figure*}

The perturbation equations of the original nonlocal version are equivalent 
to these localized equations as long as the initial conditions for $\delta X$ 
and $\delta U$ are 
\be
\mbox{IC0}: ~ \delta X(z_i) = 0\;, ~ \delta U(z_i) = 0\;, ~\delta X'(z_i) 
= 0 \;,  ~ \delta U'(z_i) = 0  \; . \label{nonlocal-IC}
\ee
Here the $z_i = 9$ is the redshift corresponding to the initial time 
$t_i = 0.55~{\rm Gyrs}$. We denote this set of initial conditions by ``IC0''.
Figure~\ref{fig:ic0-solutions} presents the various results. Note the absence
of large fluctuations in $\Phi(t,\vec{k})$, $\Psi(t,\vec{k})$ and
$\delta(t,\vec{k})$. 

\subsection{Perturbations with the ghost}

\begin{figure*}[htbp] 
\begin{center}
 \begin{tabular}{cc} 
  \includegraphics[width=0.45\textwidth]{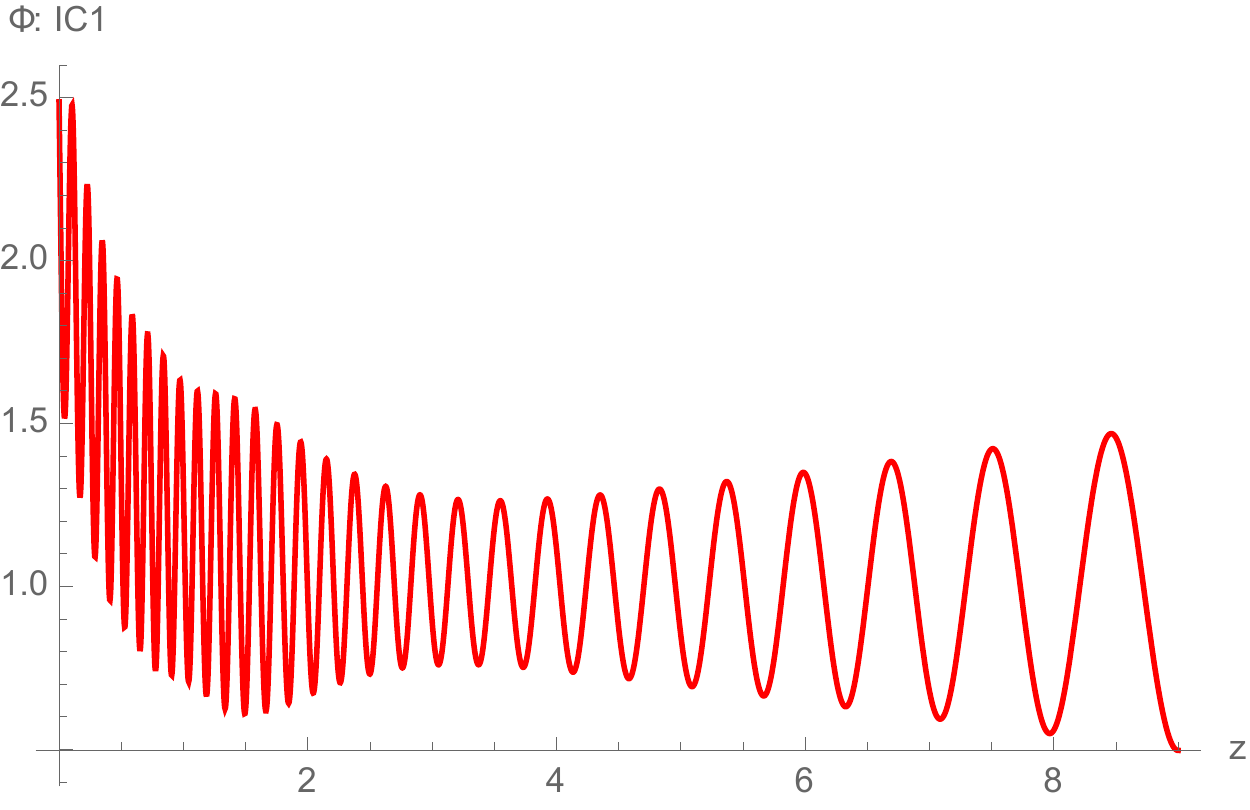}
  &
  \includegraphics[width=0.45\textwidth]{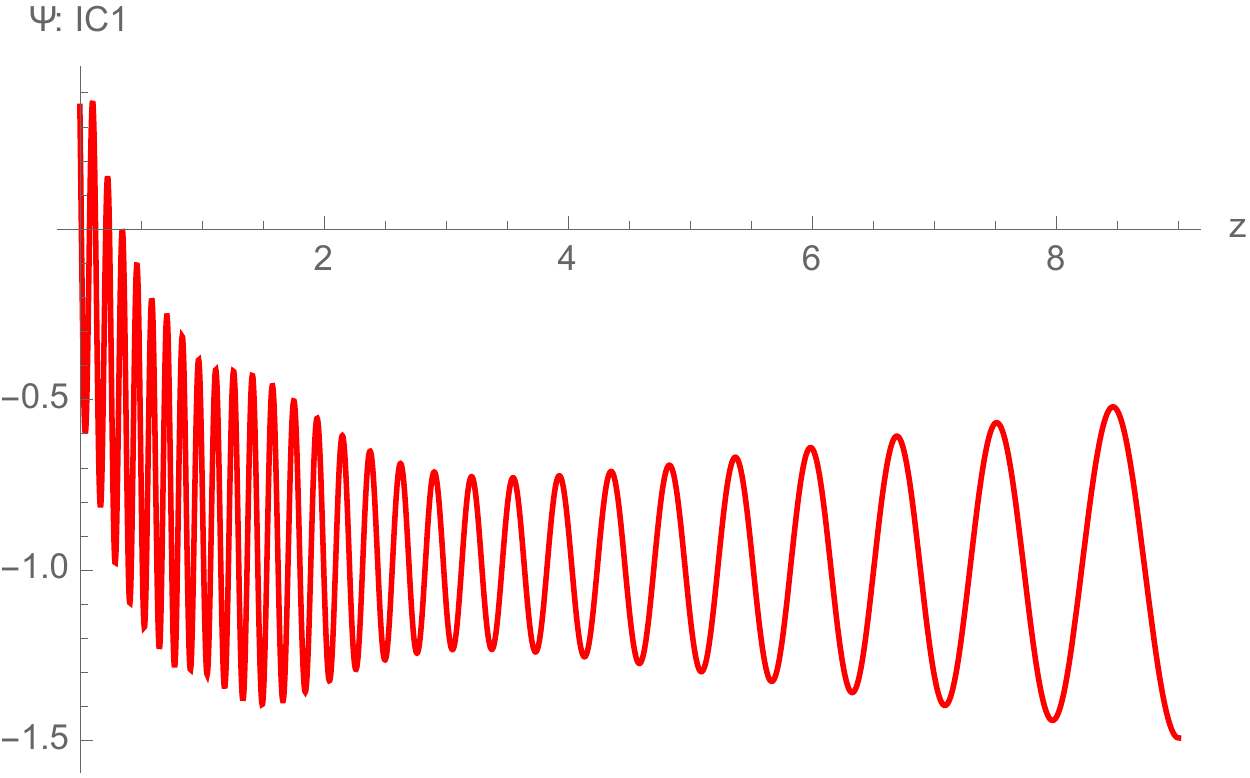}
\end{tabular}
\end{center}
\begin{center}
 \begin{tabular}{c} 
  \includegraphics[width=0.45\textwidth]{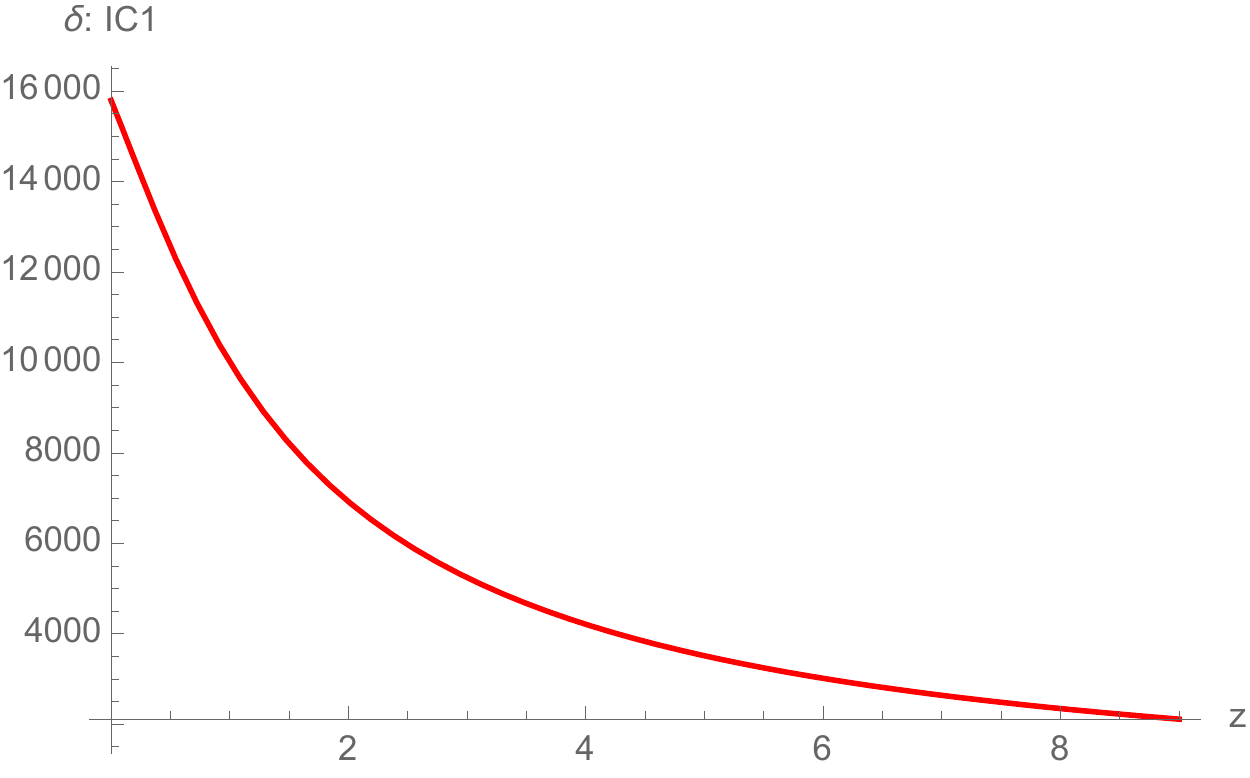} 
 \end{tabular}
\end{center}
\begin{center}
 \begin{tabular}{cc} 
  \includegraphics[width=0.45\textwidth]{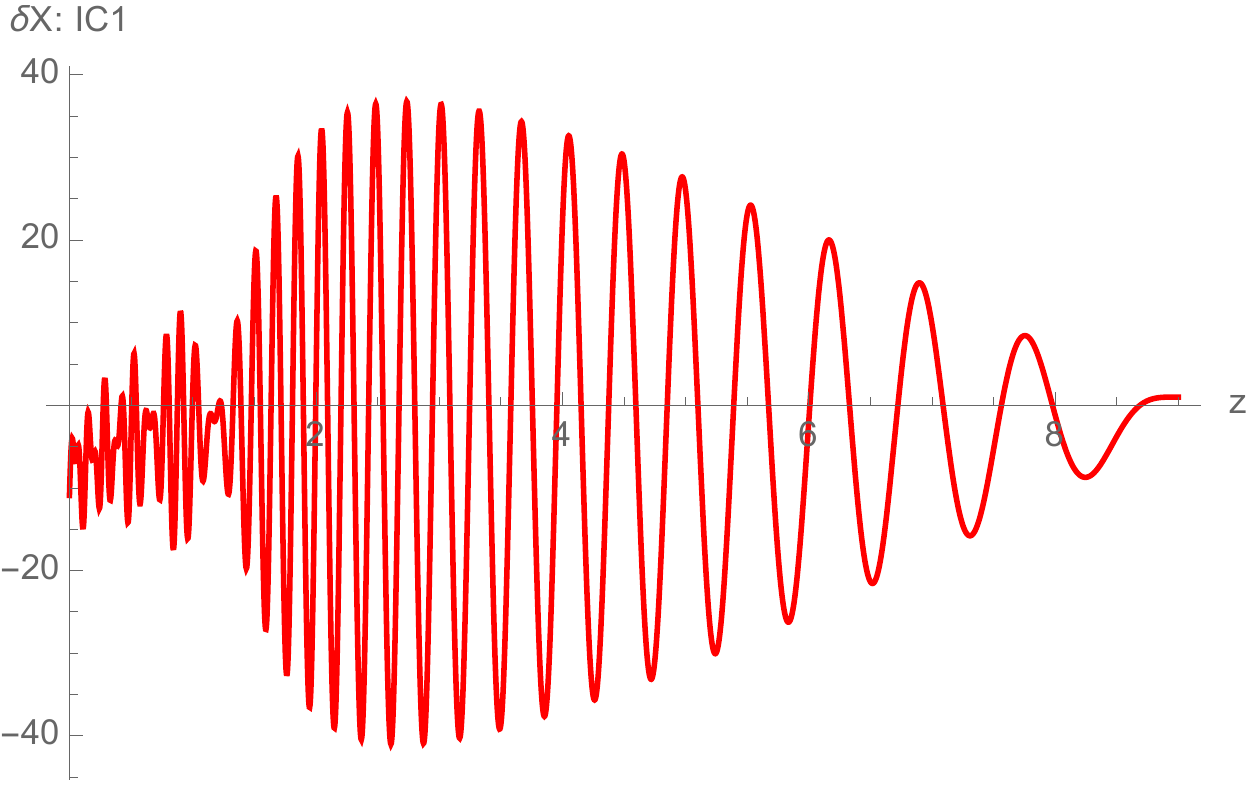}
  &
  \includegraphics[width=0.45\textwidth]{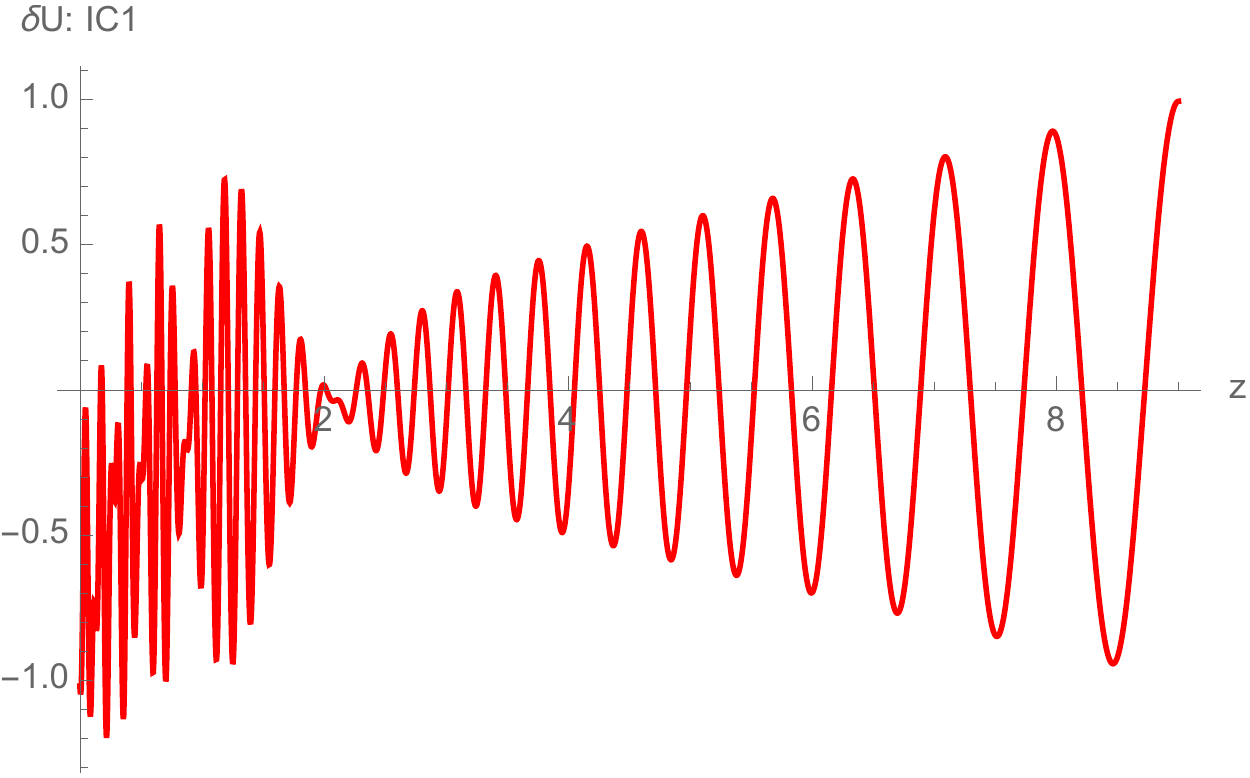}
 \end{tabular}
\end{center}
\caption{The evolution (as a function of redshift) of $\Phi(t,\vec{k})$, 
$\Psi(t,\vec{k})$, $\delta(t,\vec{k})$, $\delta X(t,\vec{k})$ and 
$\delta U(t,\vec{k})$ starting from IC1 initial conditions 
(\ref{IC1}).} 
\label{fig:ic1-solutions}
\end{figure*}

Of course there are infinitely many variations of the retarded boundary 
conditions IC0 (\ref{nonlocal-IC}). In order not to prejudice the theory towards
strong growth it makes sense to parameterize initial conditions for $\delta X(t,\vec{k})$
and $\delta U(t,\vec{k})$ in terms of the metric potentials and the density perturbation.
Because the latter fields initially agree with general relativity, for which $\Psi(t,\vec{k}) 
= -\Phi(t,\vec{k})$, we are reduced to just $\Phi(t,\vec{k})$ and $\delta(t,\vec{k})$. A
simple condition based on $\Phi(t,\vec{k})$ is,
\begin{equation}
\mbox{IC1}:  ~ 
\delta X(z_i) = \Phi(z_i)\;, ~ \delta U(z_i) = \Phi(z_i)\;, ~\delta X'(z_i) =
\Phi'(z_i)\;, ~ \delta U'(z_i) = \Phi'(z_i) \;,
\label{IC1}
\end{equation}
We call this ``IC1'' and the results for it are given in 
Fig.~\ref{fig:ic1-solutions}. Note that the fluctuations in $\delta X(t,\vec{k})$
have about 20 times the amplitude of those with the IC0 initial condition of 
Fig.~\ref{fig:ic0-solutions}. Note also that these fluctuations are communicated
to the metric potentials $\Phi(t,\vec{k})$ and $\Psi(t,\vec{k})$.

\begin{figure*}[htbp] 
\begin{center}
 \begin{tabular}{cc} 
  \includegraphics[width=0.45\textwidth]{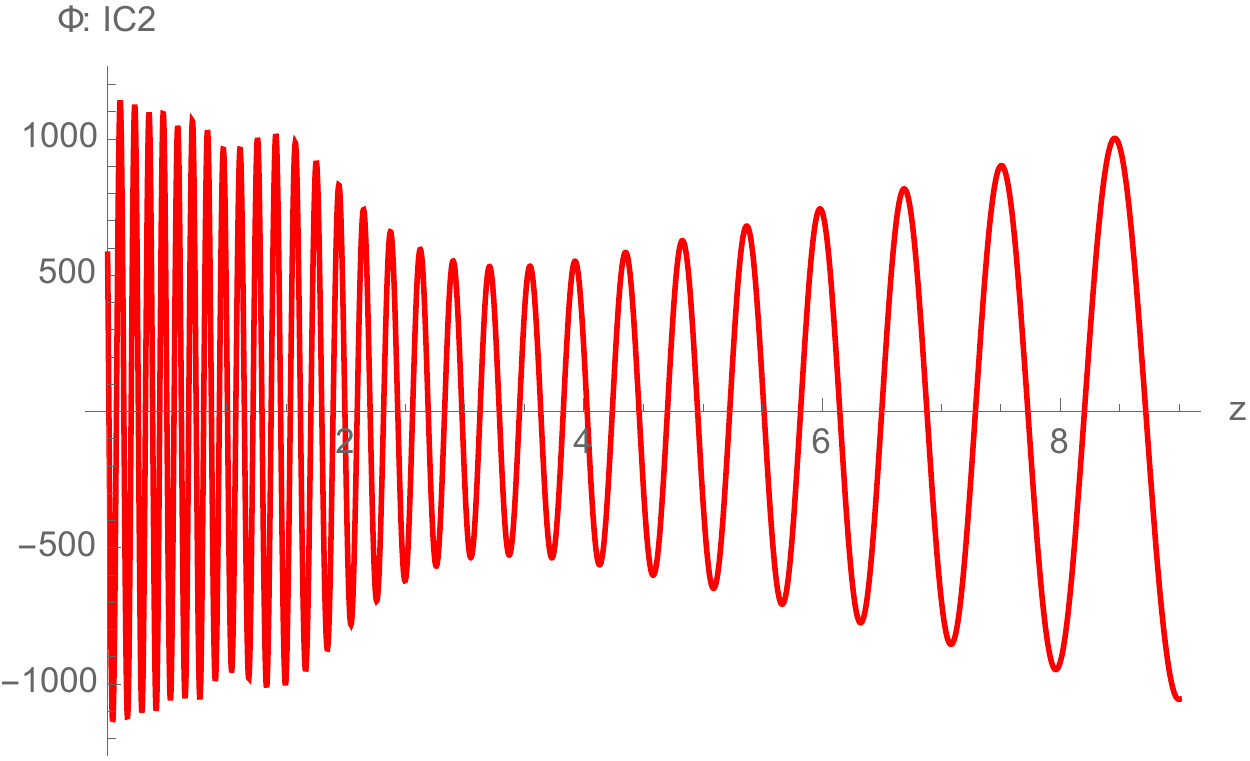}
  &
  \includegraphics[width=0.45\textwidth]{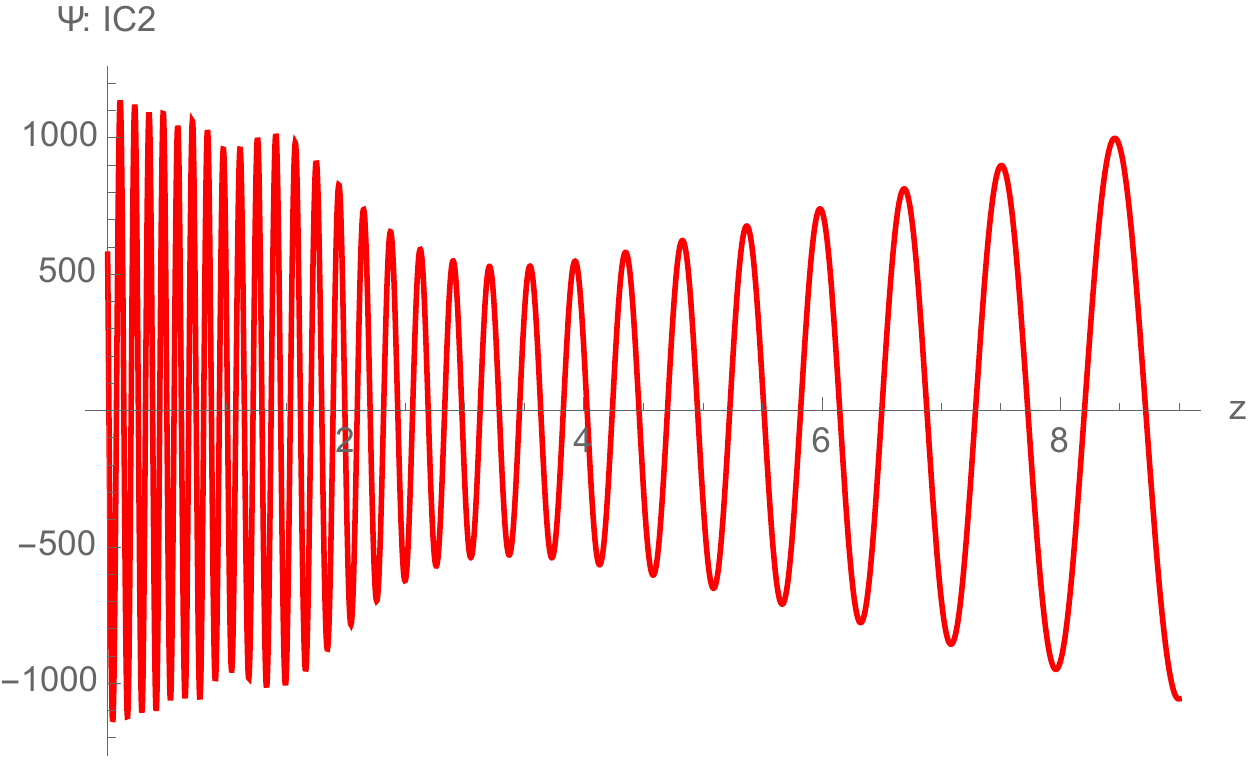}
\end{tabular}
\end{center}
\begin{center}
 \begin{tabular}{c} 
  \includegraphics[width=0.45\textwidth]{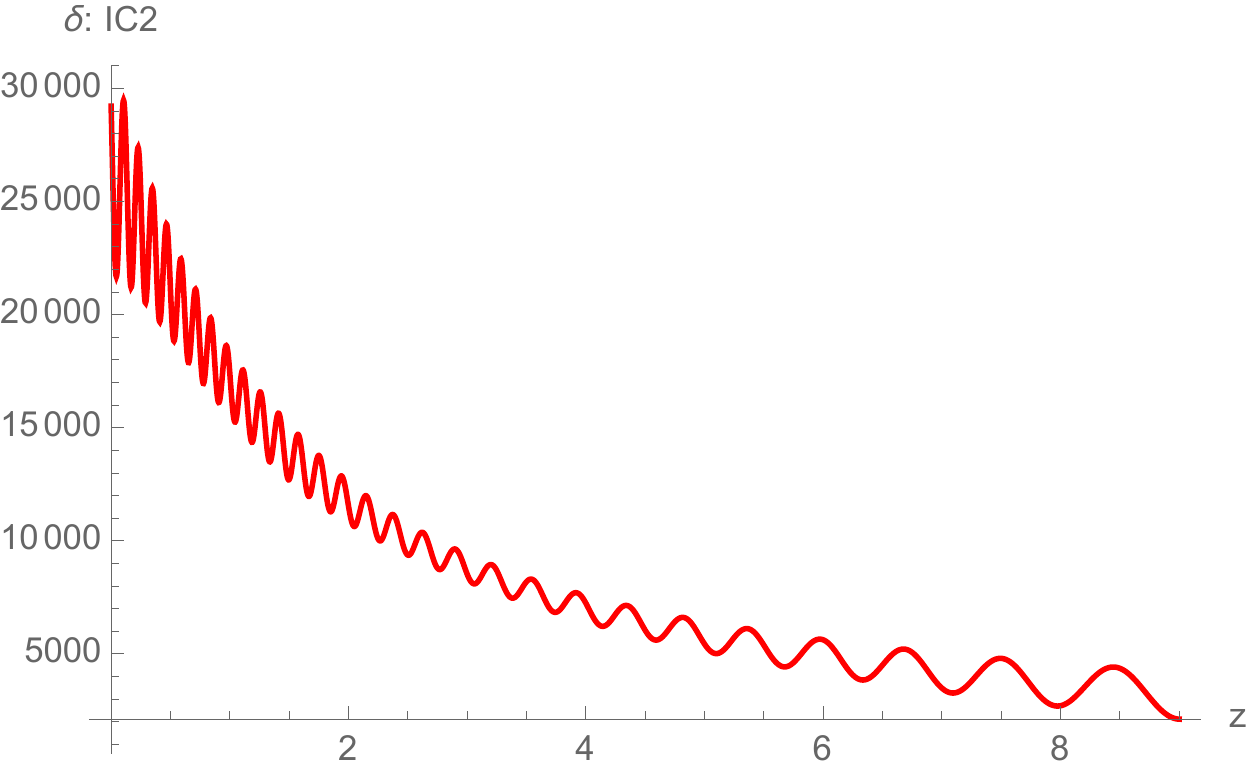} 
 \end{tabular}
\end{center}
\begin{center}
 \begin{tabular}{cc} 
  \includegraphics[width=0.45\textwidth]{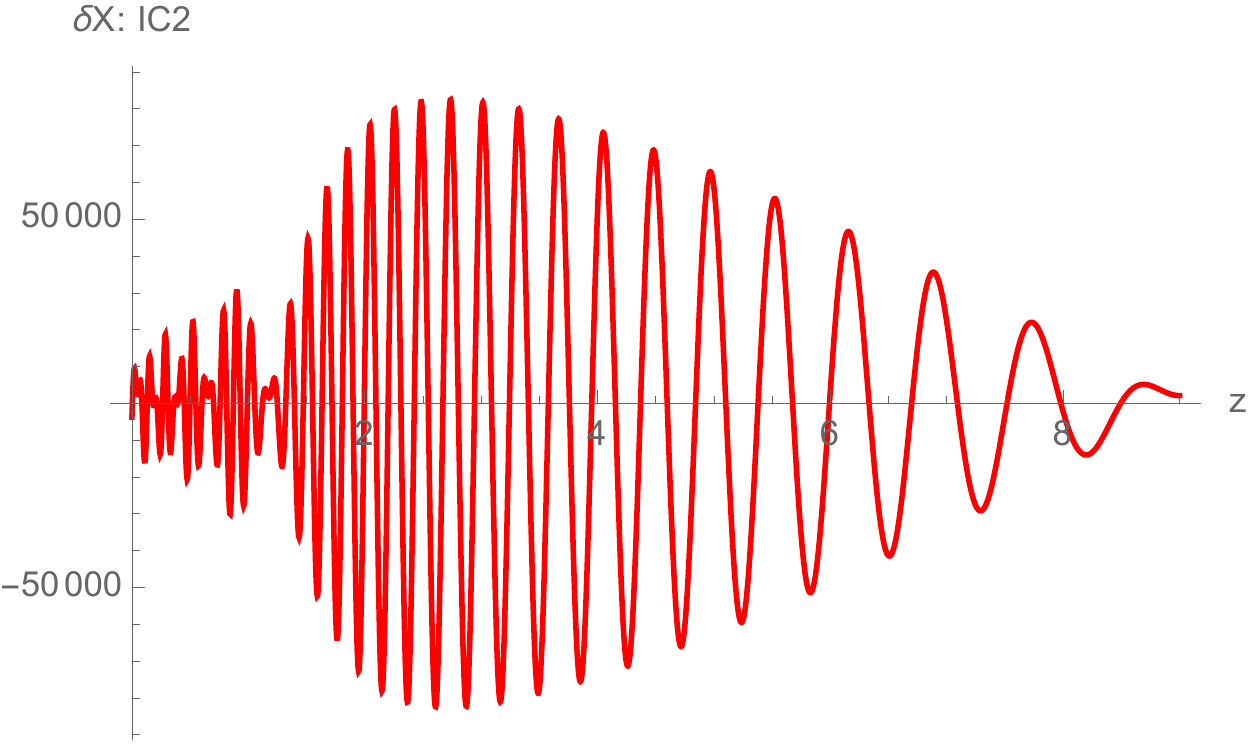}
  &
  \includegraphics[width=0.45\textwidth]{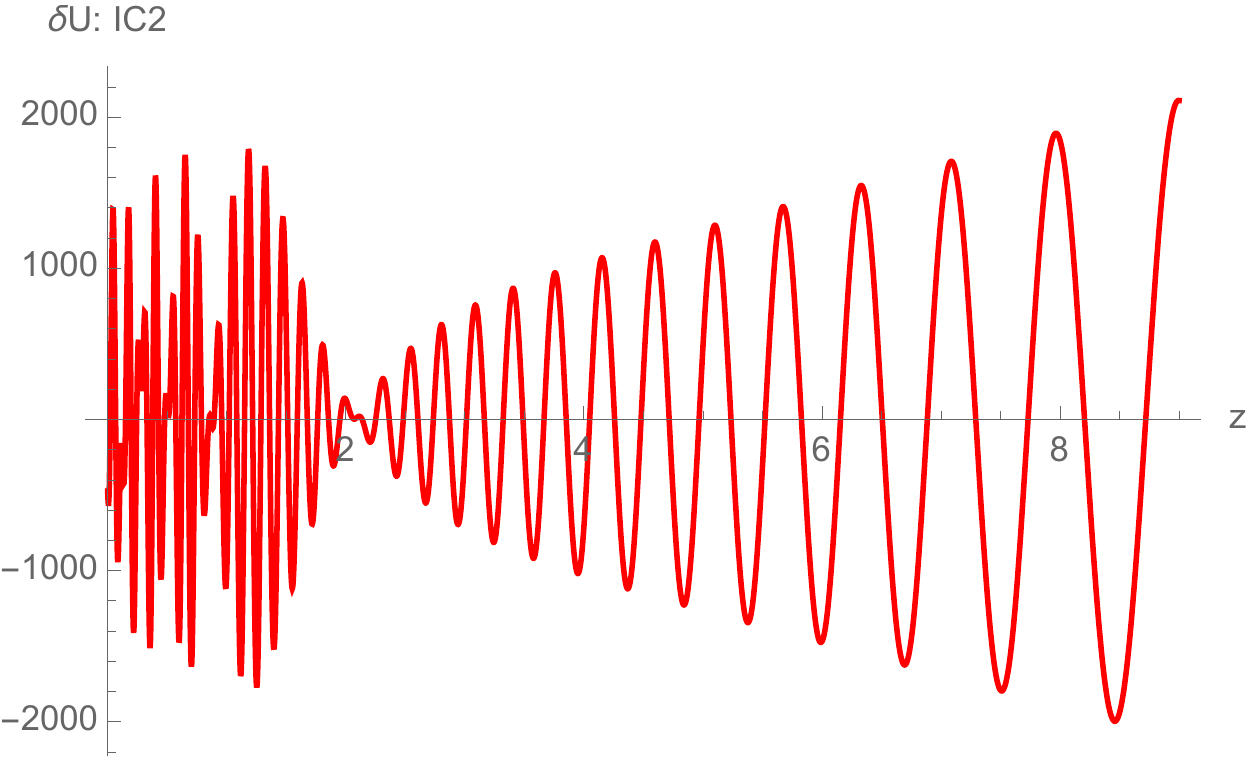}
 \end{tabular}
\end{center}
\caption{The evolution (as a function of redshift) of $\Phi(t,\vec{k})$, 
$\Psi(t,\vec{k})$, $\delta(t,\vec{k})$, $\delta X(t,\vec{k})$ and 
$\delta U(t,\vec{k})$ starting from IC2 initial conditions 
(\ref{IC2}).}
\label{fig:ic2-solutions}
\end{figure*}

A reasonable initial condition involving the density perturbation 
$\delta(t,\vec{k})$ is,
\begin{equation}
\mbox{IC2}:  ~ \delta X(z_i) = \delta(z_i)\;, ~ \delta U(z_i) = \delta(z_i) 
\;, ~\delta X'(z_i) =\delta'(z_i) \;, ~ \delta U'(z_i) = \delta'(z_i)  \;.  
\label{IC2}
\end{equation}
We call this ``IC2'' and the results for it are given in 
Fig.~\ref{fig:ic2-solutions}. Because the density perturbation is so much 
larger than the metric potentials the resulting fluctuations in $\delta 
X(t,\vec{k})$ have about $40,000$ times the amplitude of those with the 
IC0 initial condition of Fig.~\ref{fig:ic0-solutions}! The fluctuations of 
the metric potentials $\Phi(t,\vec{k})$ and $\Psi(t,\vec{k})$ are similarly
enhanced with respect to those of IC0.

We explored many other initial conditions, for example,
\bea
&&\mbox{IC1-a}:  ~ 
\delta X(z_i) = \Phi(z_i)\;, ~ \delta U(z_i) = \Phi(z_i) \; , 
~\delta X'(z_i) =-\Phi'(z_i)\;, ~ \delta U'(z_i) = -\Phi'(z_i) \;,  
\label{IC1-a}
\\
&&\mbox{IC1-b}:  ~ 
\delta X(z_i) = \Phi(z_i)\;, ~ \delta U(z_i) = \Phi(z_i) \; , 
~\delta X'(z_i) =\Phi(z_i)\;, ~ \delta U'(z_i) = \Phi(z_i) \;.  
\label{IC1-b}
\eea
We have not reported them because the results are very similar to those
of IC1 (\ref{IC1}). The same is true for the variants of IC2 (\ref{IC2}) 
which involve the density perturbation $\delta(t,\vec{k})$.

\subsection{Comparing IC0 with IC1 and IC2}

\begin{figure*}[htbp] 
\begin{center}
 \begin{tabular}{cc} 
  \includegraphics[width=0.45\textwidth]{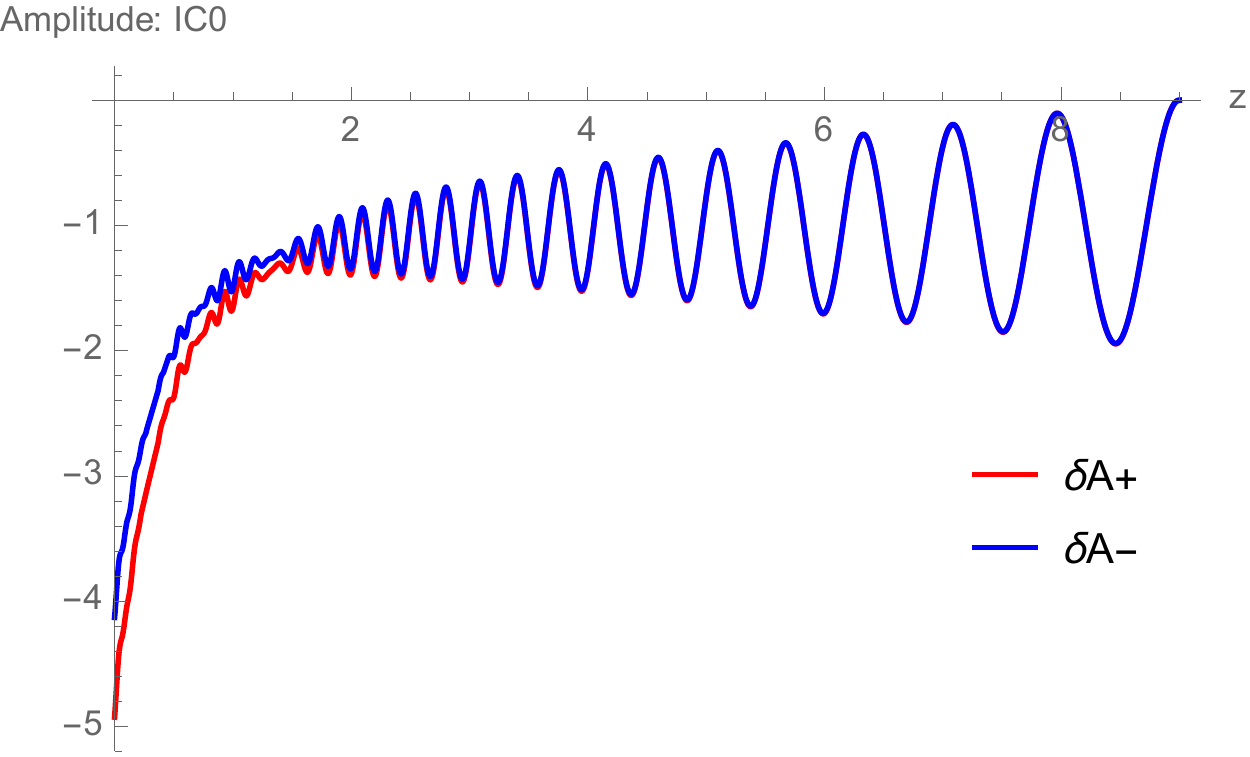}
  &
  \includegraphics[width=0.45\textwidth]{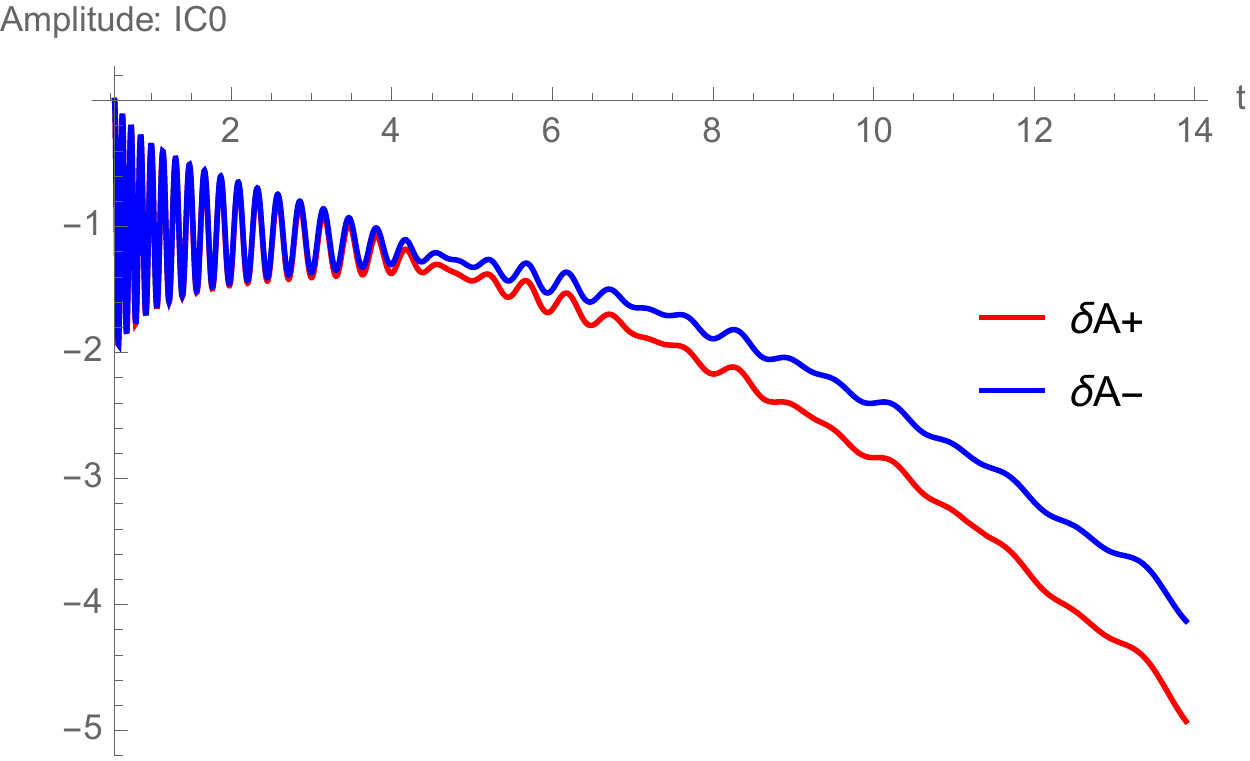}
\end{tabular}
\end{center}
\begin{center}
 \begin{tabular}{cc} 
  \includegraphics[width=0.45\textwidth]{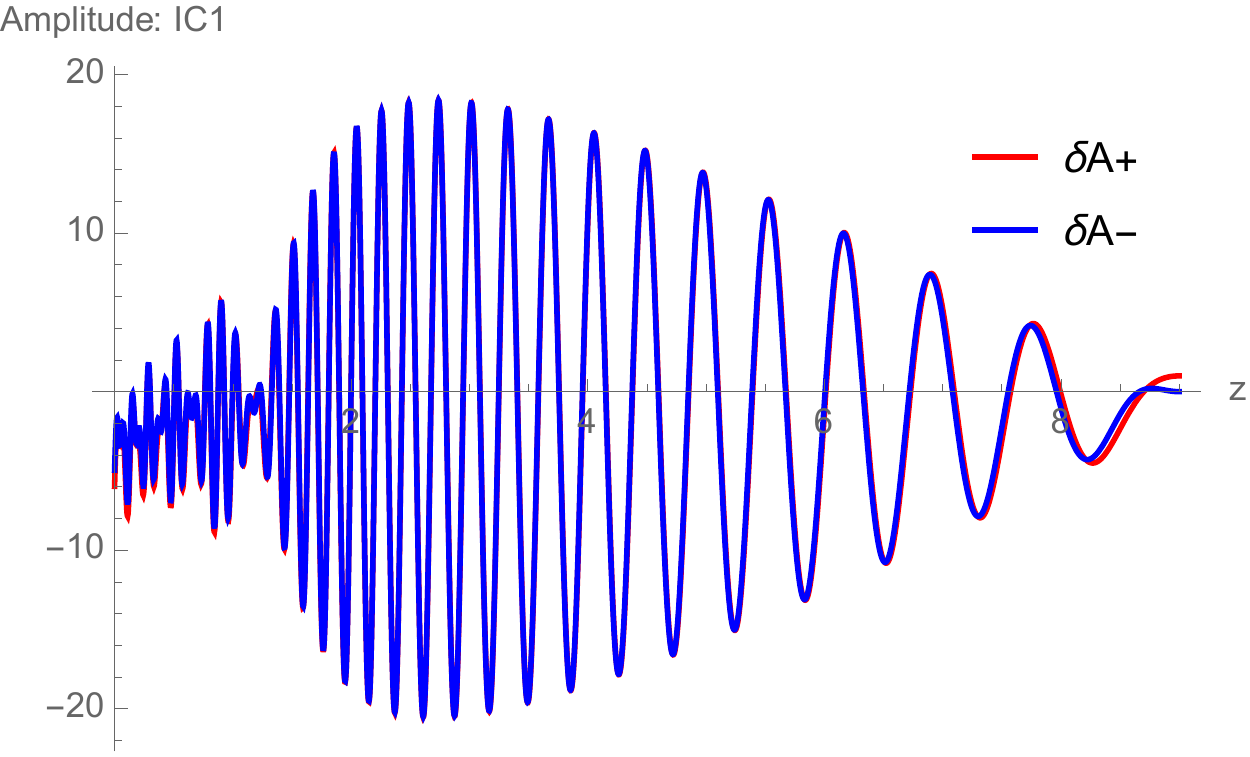} 
  &
  \includegraphics[width=0.45\textwidth]{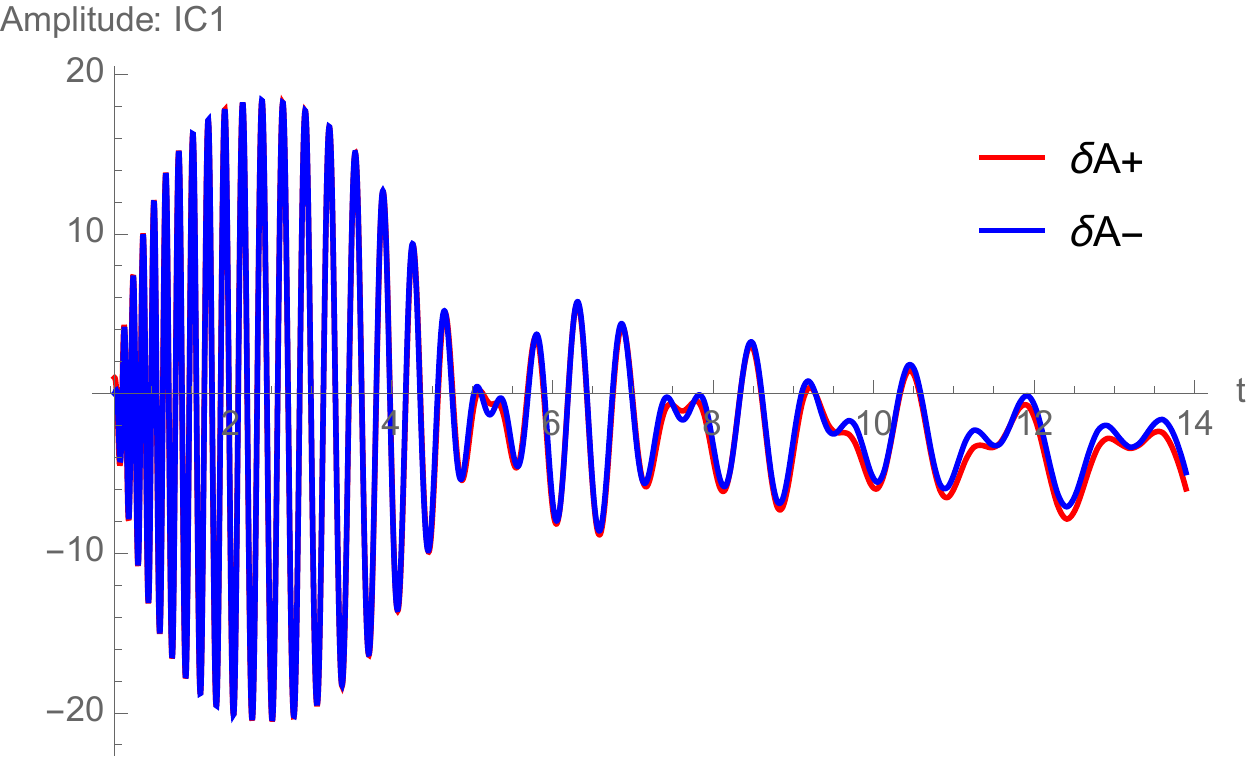} 
 \end{tabular}
\end{center}
\begin{center}
 \begin{tabular}{cc} 
  \includegraphics[width=0.45\textwidth]{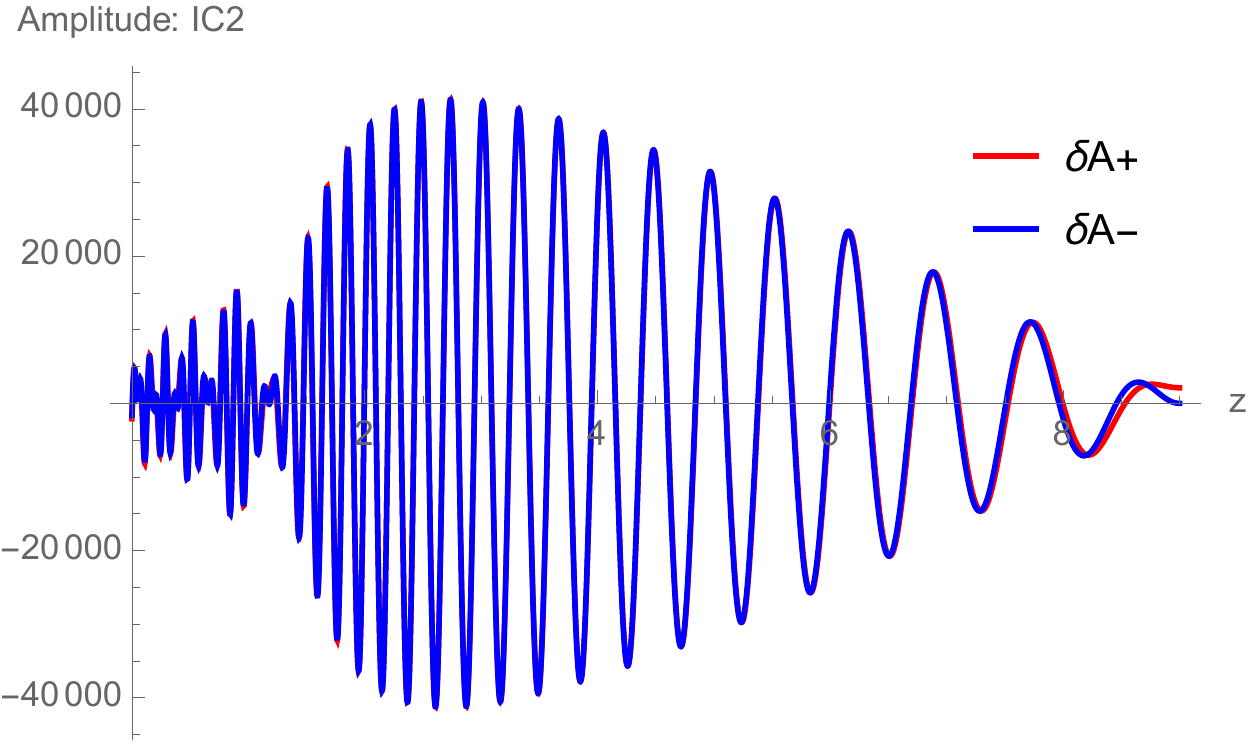}
  &
  \includegraphics[width=0.45\textwidth]{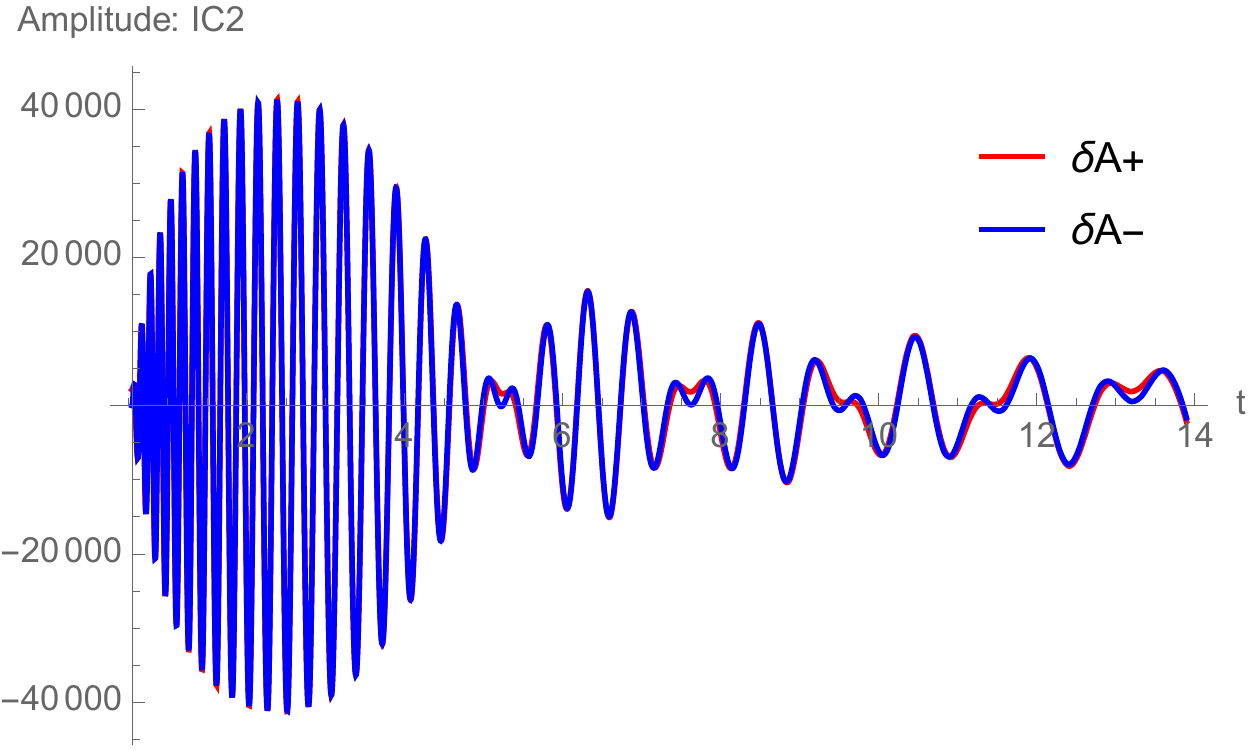}
 \end{tabular}
\end{center}
\caption{Amplitudes of the perturbations $\delta A_{\pm} = \frac12 
(\delta X \pm \delta U)$ versus the redshift $z$ (on the left) and 
versus the co-moving time $t$ (in the right) for the initial conditions
IC0 (\ref{nonlocal-IC}), IC1 (\ref{IC1}), and IC2 (\ref{IC2}).} 
\label{fig:amplitude-z-t}
\end{figure*}

Figure~\ref{fig:ic0-solutions} shows the evolution of normal perturbations,
whereas Figures~\ref{fig:ic1-solutions} and \ref{fig:ic2-solutions} depict
the evolution of perturbations in which the ghost field $\delta A_+ = \frac12 [
\delta X(t,\vec{k}) + \delta U(t,\vec{k})]$ is excited. The contrast between
IC0 (without the ghost) and the other conditions IC1-2 (with the ghost) is
striking. It becomes even more so in Figure~\ref{fig:amplitude-z-t}, which
displays just $\delta A_{\pm}$ for all three cases, both as functions of redshift
$z$ and as functions of the co-moving time $t$. Recall that a kinetic
instability manifests through the ghost field ($\delta A_+$) experiencing a wild
time evolution, and conserving energy by dragging along the normal fields,
in this case $\delta A_-$. The IC1 and IC2 initial conditions show this quite 
clearly, whereas that behavior is not at all apparent with the IC0 initial 
conditions.

\begin{figure*}[htbp]
\begin{center}
 \begin{tabular}{cc} 
  \includegraphics[width=0.45\textwidth]{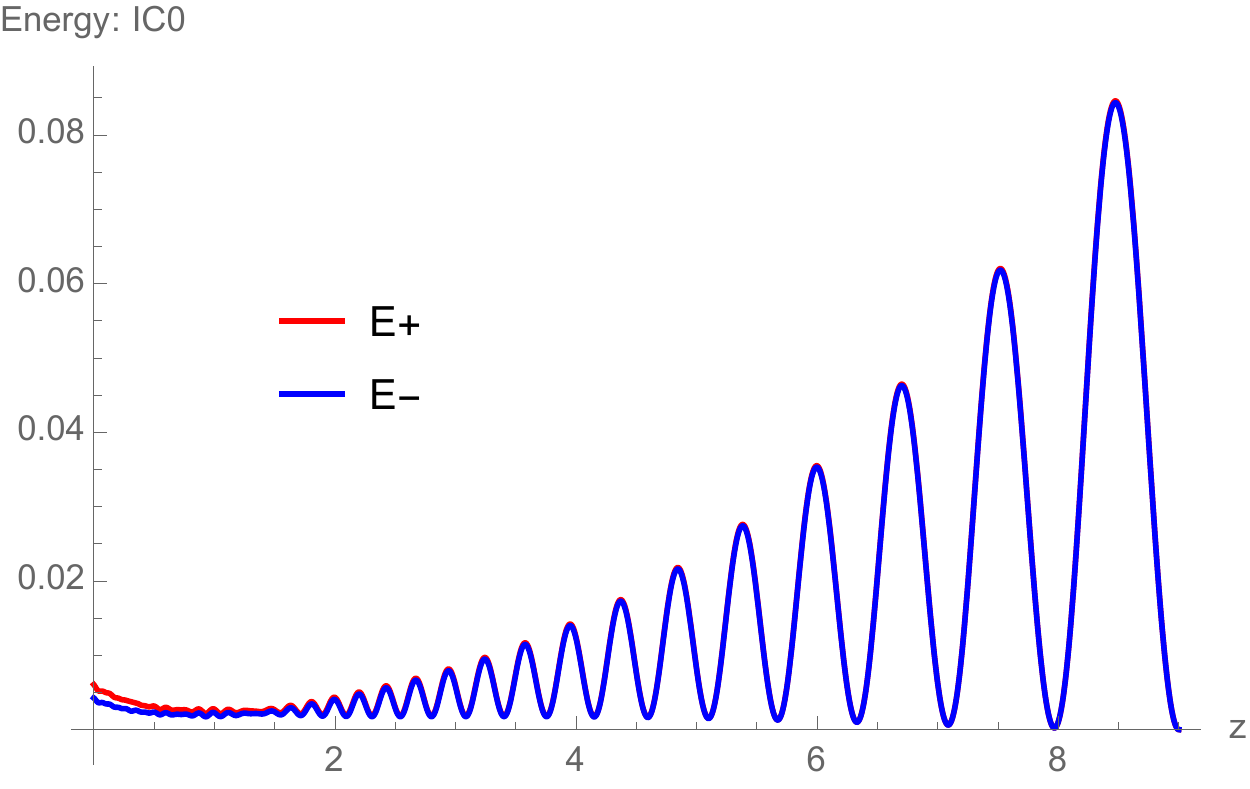}
  &
  \includegraphics[width=0.45\textwidth]{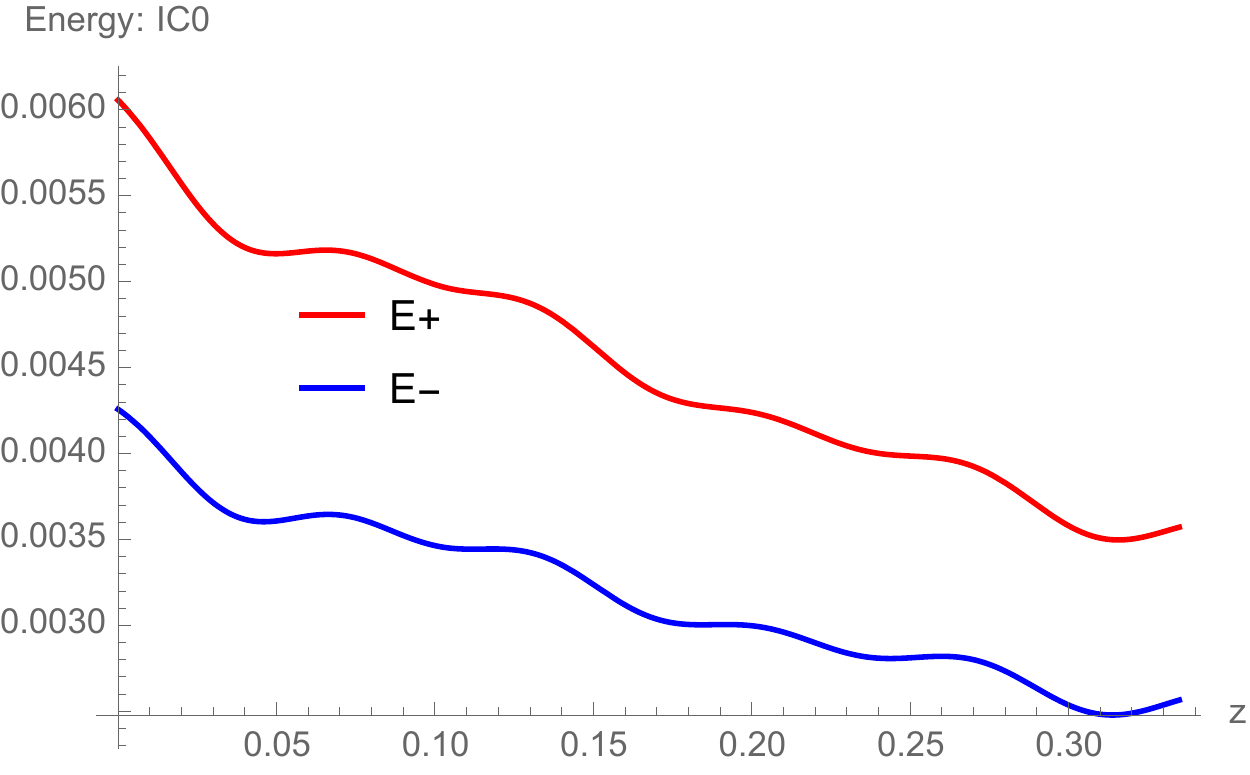}
\end{tabular}
\end{center}
\begin{center}
 \begin{tabular}{cc} 
  \includegraphics[width=0.45\textwidth]{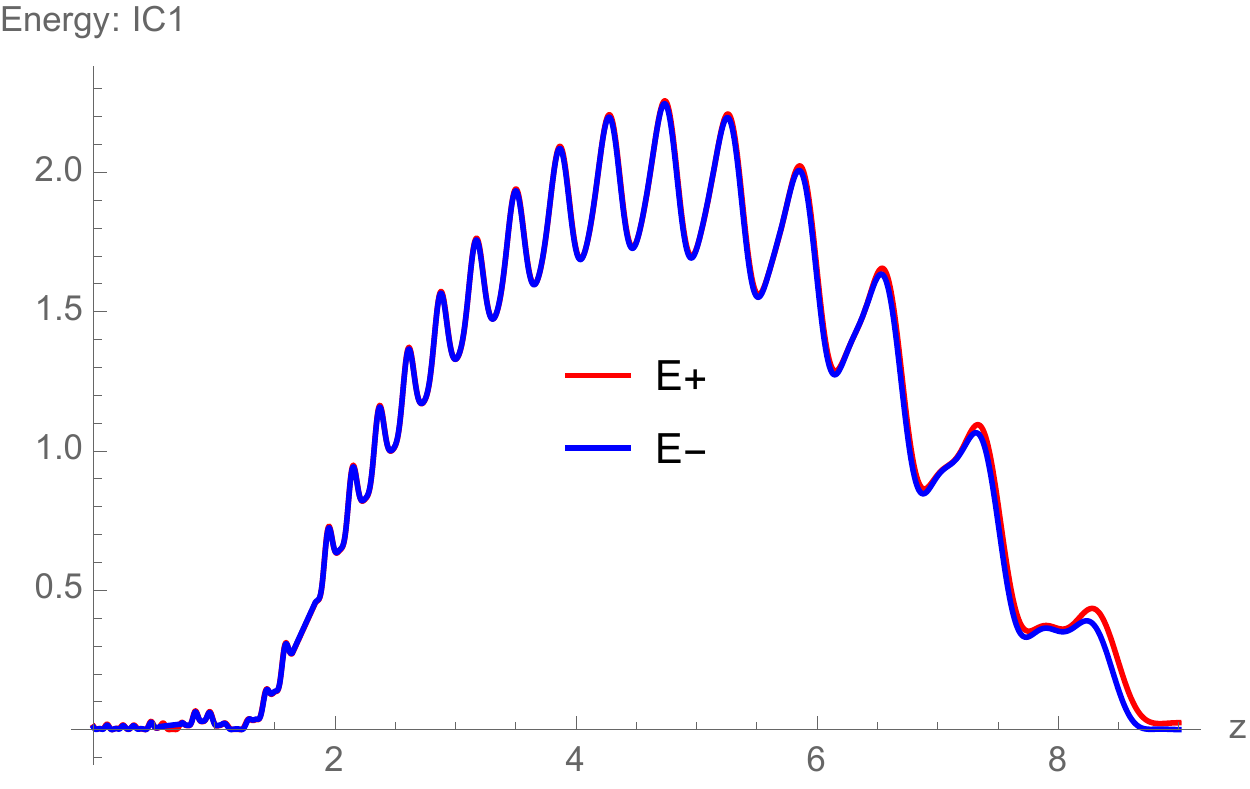} 
  &
  \includegraphics[width=0.45\textwidth]{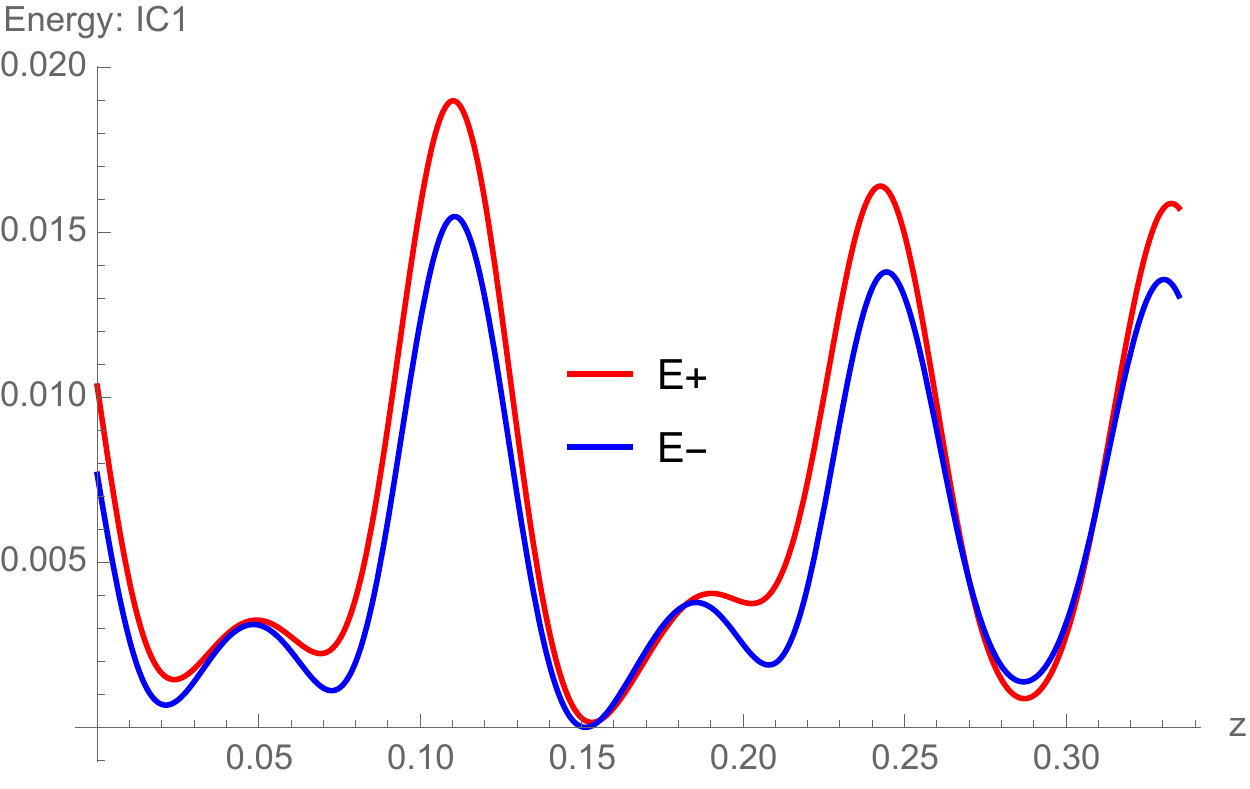} 
 \end{tabular}
\end{center}
\begin{center}
 \begin{tabular}{cc} 
  \includegraphics[width=0.45\textwidth]{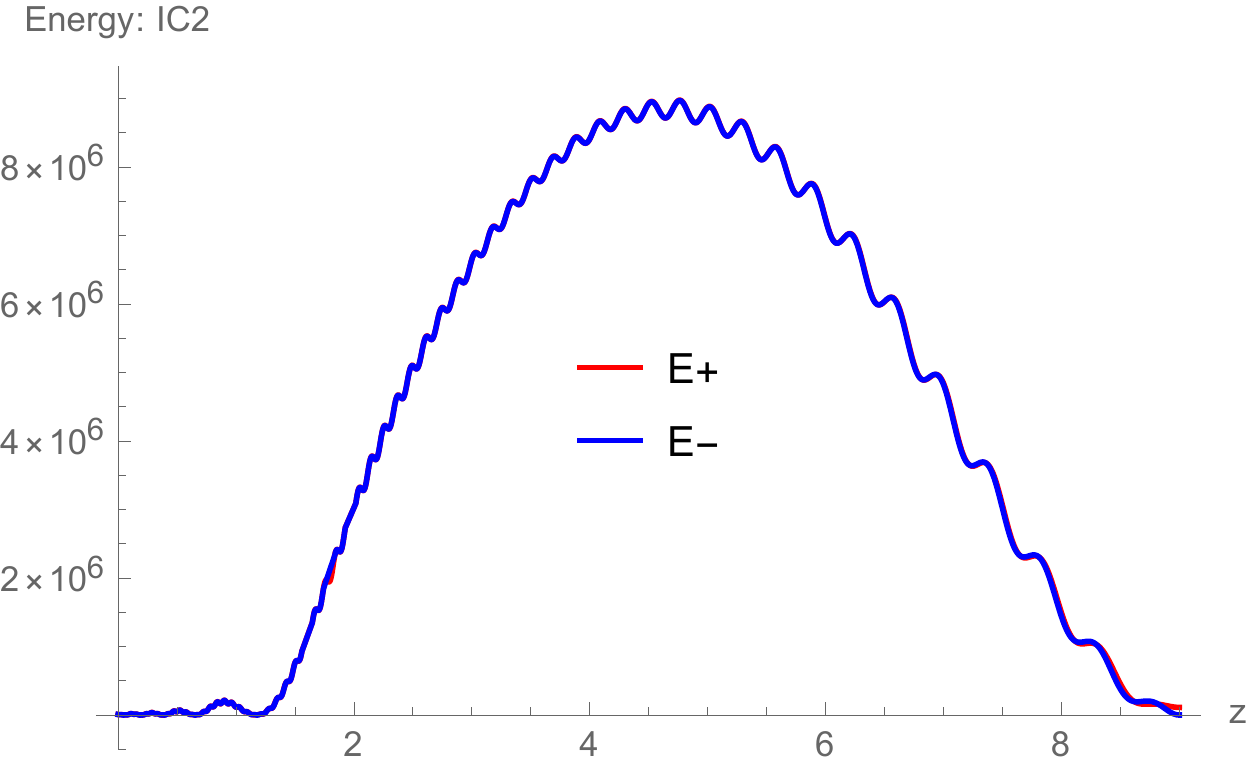}
  &
  \includegraphics[width=0.45\textwidth]{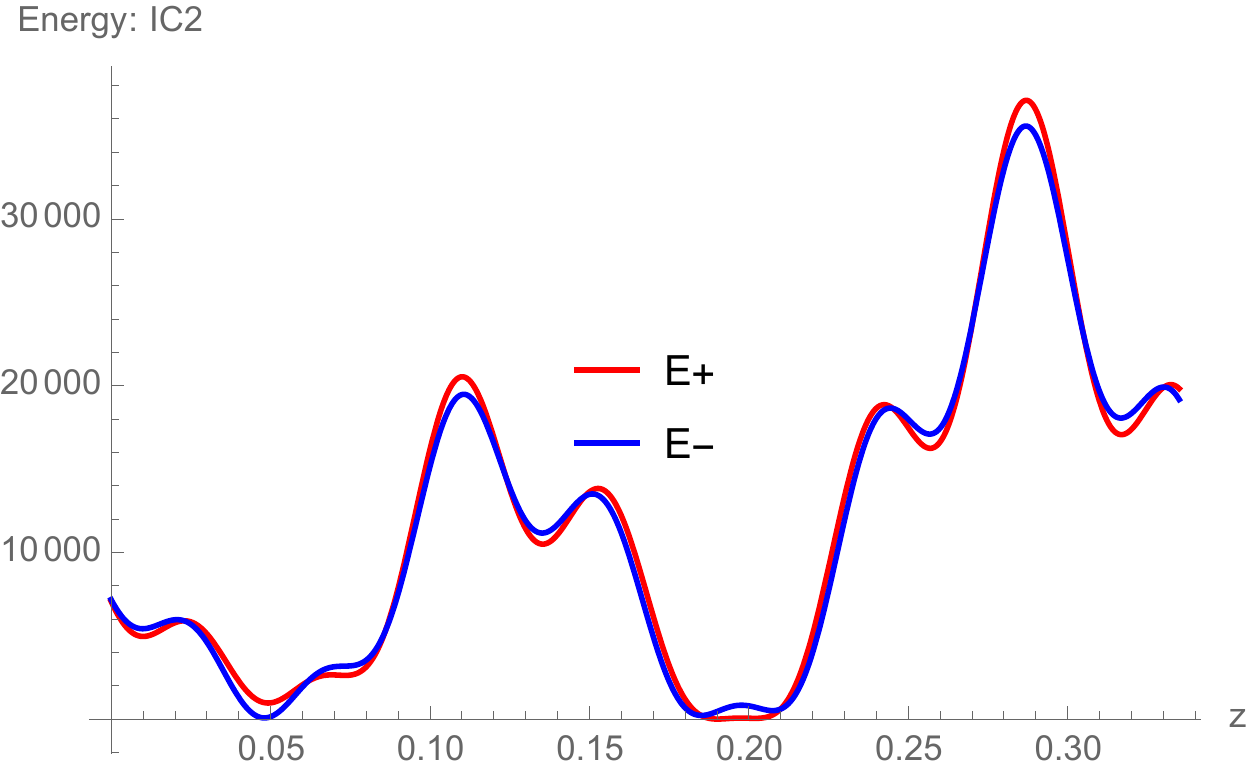}
 \end{tabular}
\end{center}
\caption{The kinetic energies $E_{\pm}$ (\ref{Epm}) versus redshift $z$
for the initial conditions IC0 (\ref{nonlocal-IC}), IC1 (\ref{IC1}), and 
IC2 (\ref{IC2}). In each case the left hand graphs show the full range
$0 < z < 9$, whereas the right hand graphs provide an expanded view of
the late time regime $0 < z < 0.34$.}
\label{fig:energy-z}
\end{figure*}

\begin{figure*}[htbp] 
\begin{center}
 \begin{tabular}{cc} 
  \includegraphics[width=0.45\textwidth]{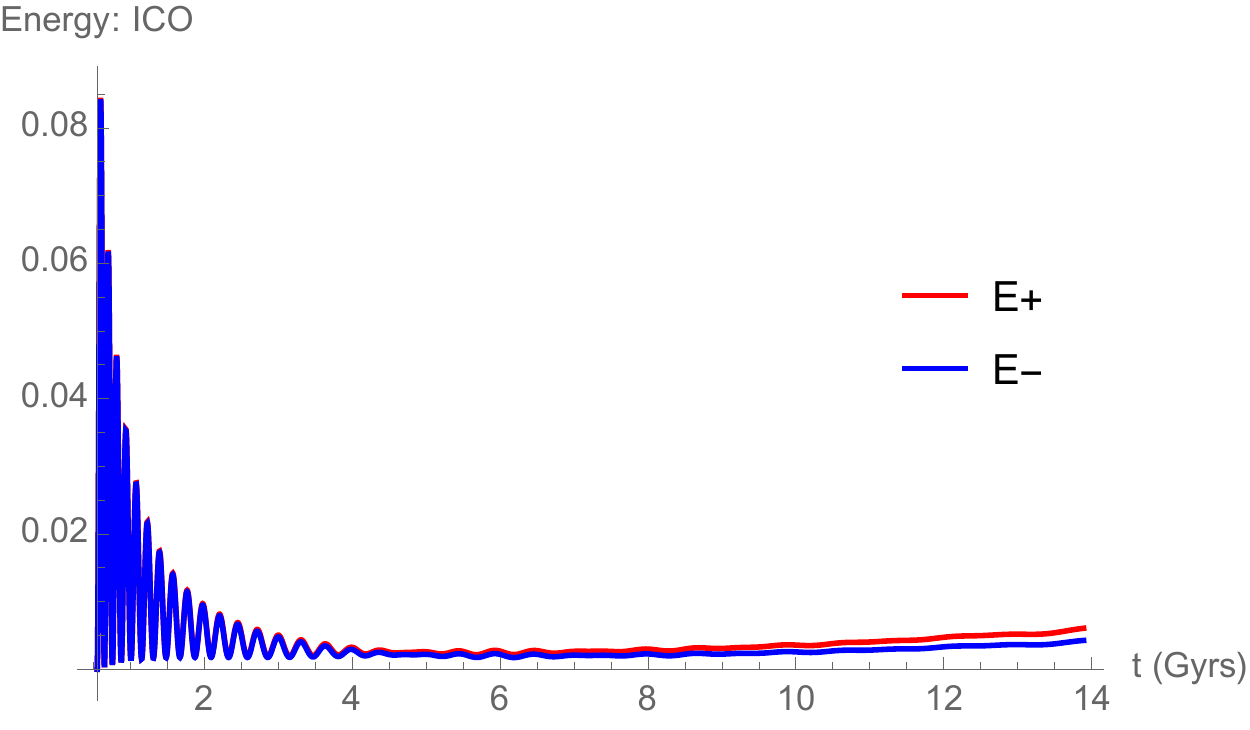}
  &
  \includegraphics[width=0.45\textwidth]{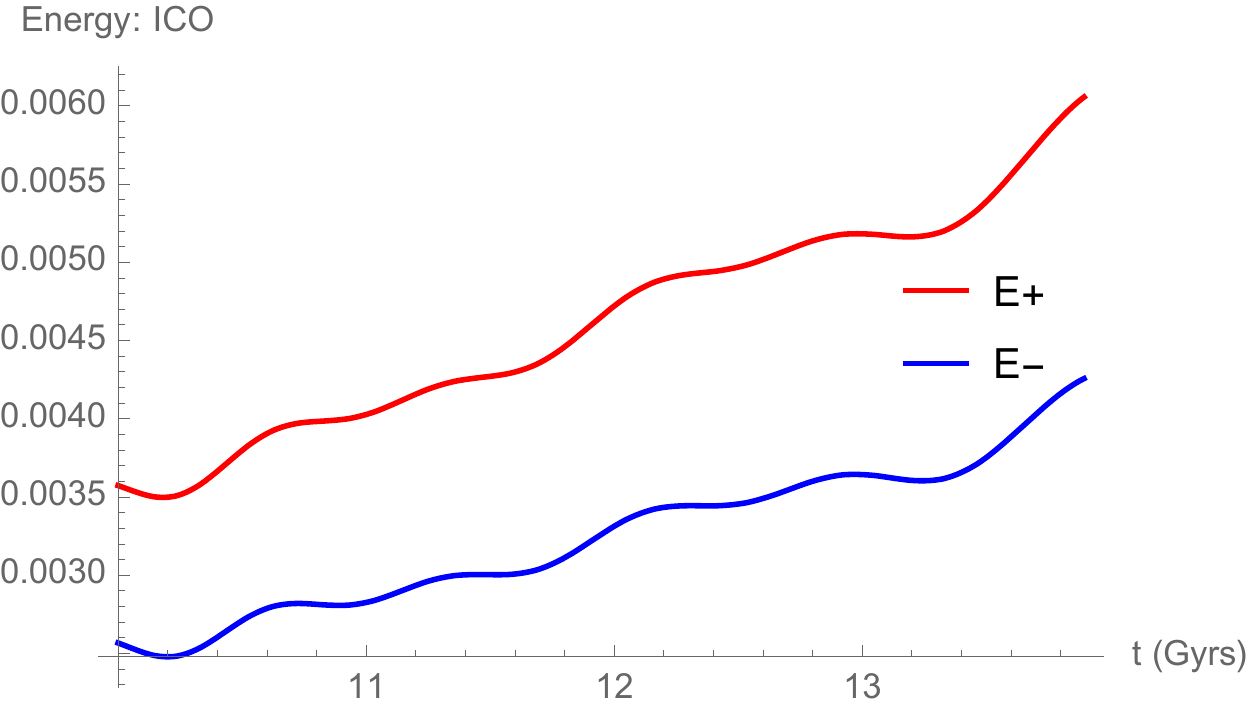}
\end{tabular}
\end{center}
\begin{center}
 \begin{tabular}{cc} 
  \includegraphics[width=0.45\textwidth]{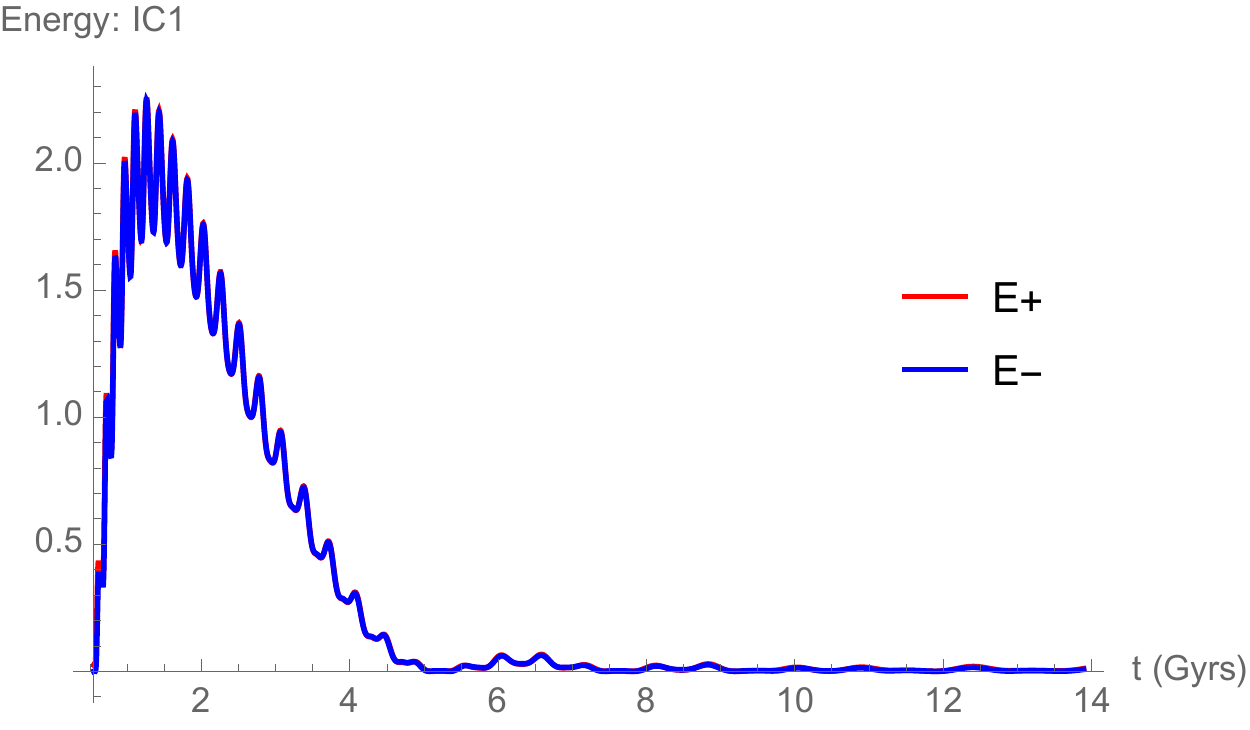} 
  &
  \includegraphics[width=0.45\textwidth]{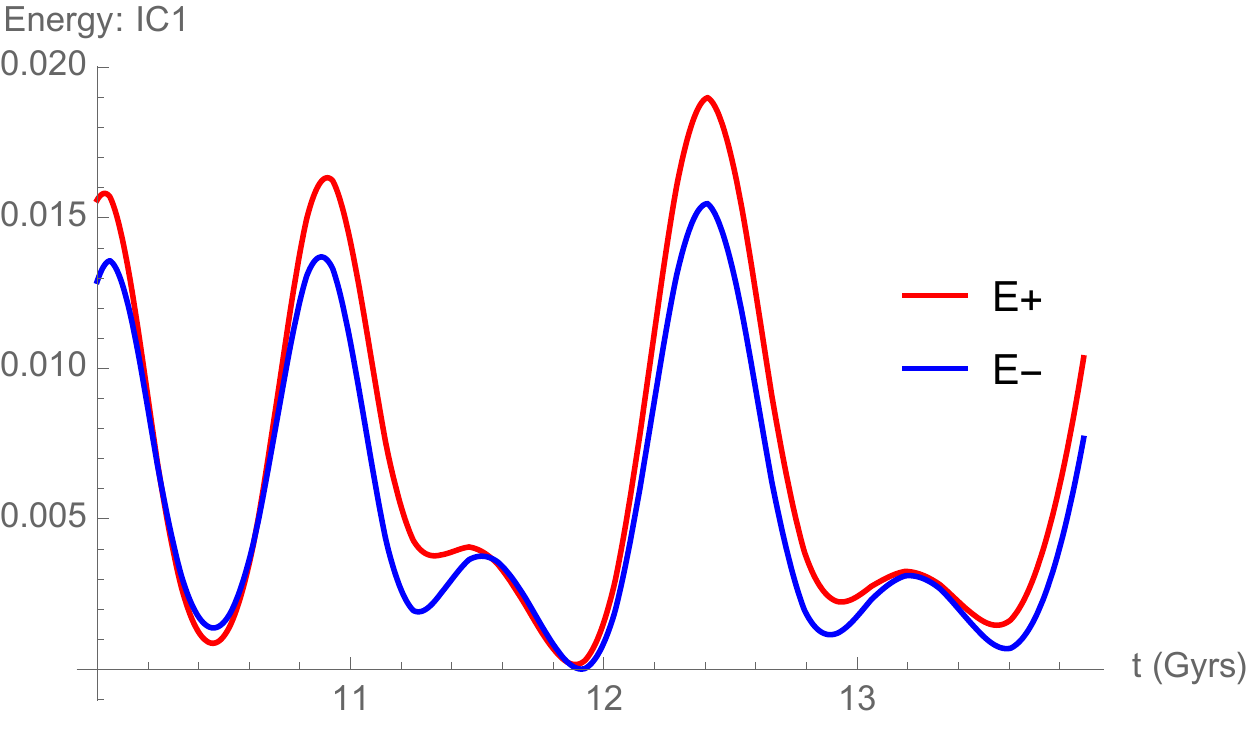} 
 \end{tabular}
\end{center}
\begin{center}
 \begin{tabular}{cc} 
  \includegraphics[width=0.45\textwidth]{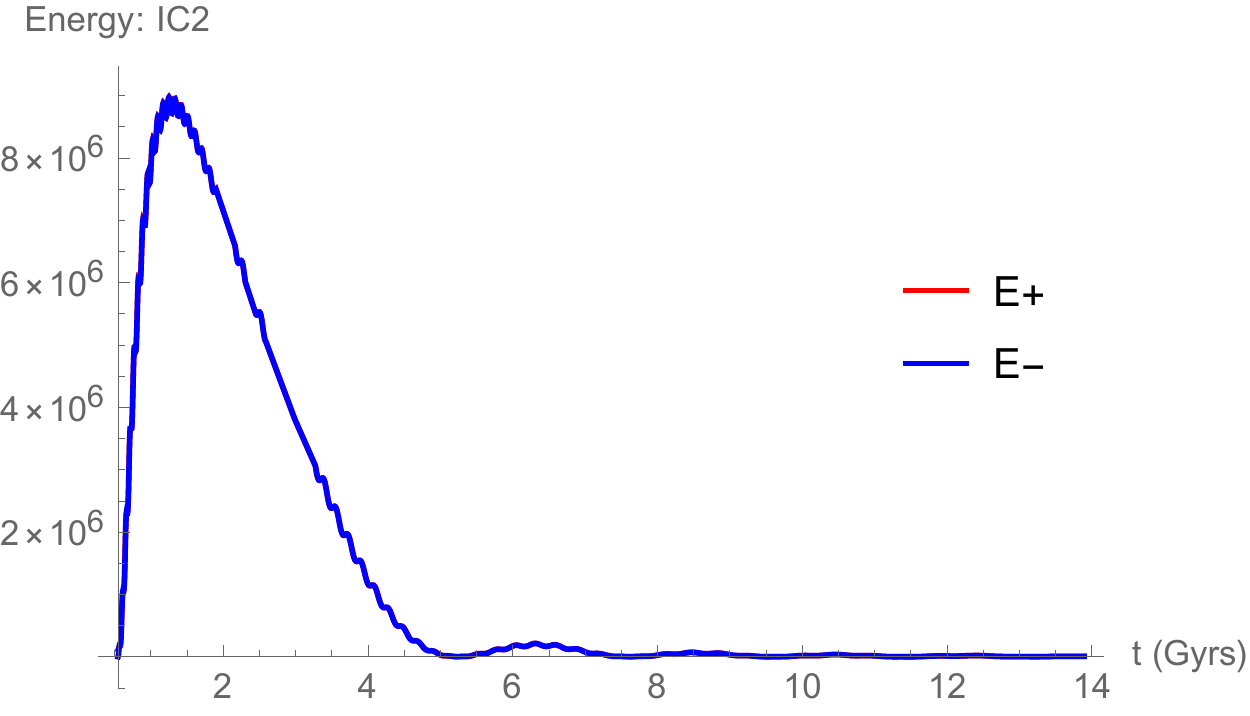}
  &
  \includegraphics[width=0.45\textwidth]{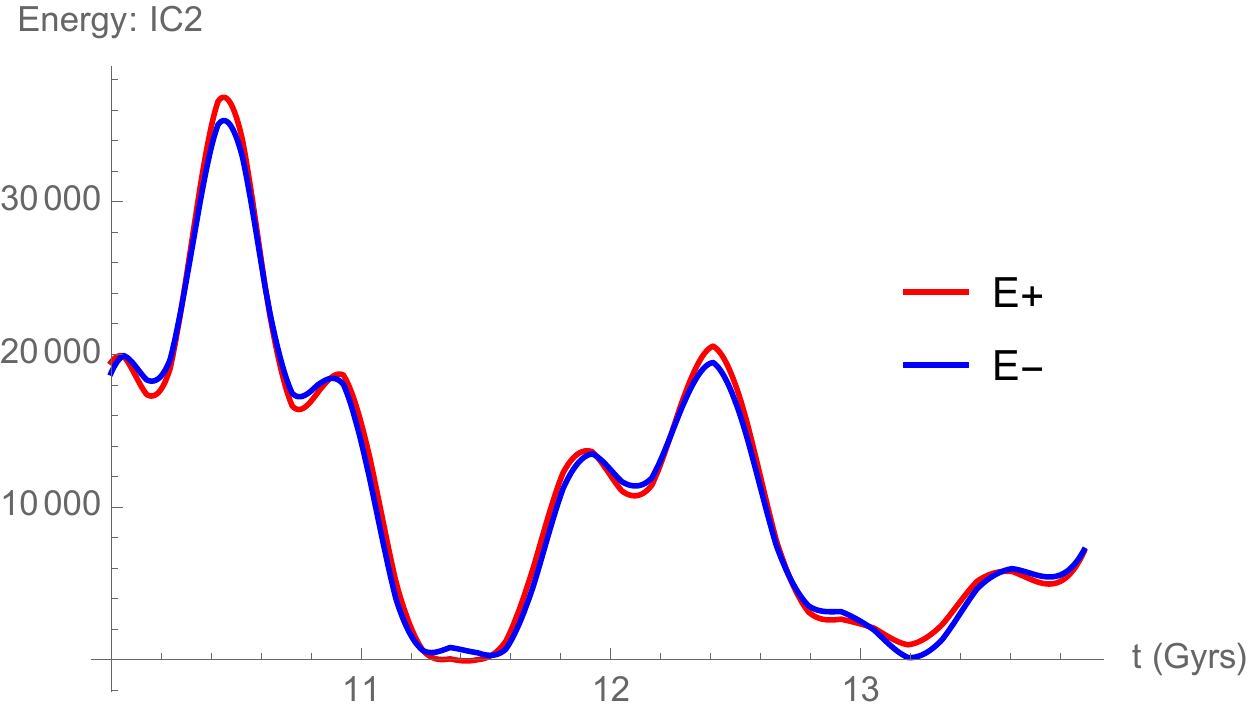}
 \end{tabular}
\end{center}
\caption{The kinetic energies $E_{\pm}$ (\ref{Epm}) versus the co-moving
time $t$ for the initial conditions IC0 (\ref{nonlocal-IC}), IC1 (\ref{IC1}), 
and IC2 (\ref{IC2}). In each case the left hand graph shows the full range
$0.55~{\rm Gyr} < t < 13.89~{\rm Gyr}$ whereas the right hand graphs provide 
an expanded view of the late time regime $10~{\rm Gyr} < t < 13.89~{\rm Gyr}$.}
\label{fig:energy-time}
\end{figure*}

Of course the amplitudes $\delta A_{\pm}$ have no immediate physical meaning.
Because the ghost instability is associated with kinetic energy we would like a
measure of how much kinetic energy resides in $A_{\pm}$. Of course there is no 
true energy functional for gravity in cosmology so one cannot expect complete 
precision but a rough measure of the kinetic energy in $A_{\pm}$ derives from
their stress tensors
\begin{equation}
\mp T^{\pm}_{\mu\nu} = \partial_{\mu} A_{\pm} \partial_{\nu} A_{\pm} - \frac12
g_{\mu\nu} g^{\rho\sigma} \partial_{\rho} A_{\pm} \partial_{\sigma} 
A_{\pm} \; . \label{stress}
\end{equation}
One can recognize $T^{+}_{\mu\nu} + T^{-}_{\mu\nu}$ as the final term in
relation (\ref{DeltaGmn-local}) for the localized version of the modified
Einstein tensor. Perturbing $T^{\pm}_{\mu\nu}$ around the cosmological 
background induces linearized spatial plane wave contributions which drop out 
of equations (\ref{modified-Poisson-eq-summary}-\ref{modified-gslip-eq-summary}) 
in the sub-horizon regime of $k \gg Ha$. At quadratic order there are diagonal
terms and mixings between the metric perturbations and the auxiliary scalars. A 
rough measure of how much kinetic energy resides in $A_{\pm}$ comes from the 
diagonal contributions,
\be
E_{\pm} \equiv \frac{1}{2}\dot{\delta A_{\pm}^2} + \frac{1}{2}
\frac{k^2}{a^2}\delta A_{\pm}^2 \;. \label{Epm}
\ee
Because the actual kinetic energy of the ghost mode is $-E_+$ we see
again the terrible instability associated with ghosts. It costs {\it
zero} total energy to start with arbitrarily large values of
$\delta A_{+}(0,\vec{k}) = \delta A_{-}(0,\vec{k})$, at arbitrarily
large wave numbers. That is all precluded by the retarded initial
conditions of the original, nonlocal theory (\ref{DW-action}) but it
is a fatal problem for the localized model (\ref{DW-action-localized-IBP}).

Figures~\ref{fig:energy-z} and \ref{fig:energy-time} show $E_{\pm}$ for
each of the three initial conditions, first as functions of the redshift $z$
and then in terms of the co-moving time $t$. For the non-ghost initial
condition IC0 the energies $E_+$ and $E_-$ have distinct evolutions, and 
are quite small. In contrast, $E_{\pm}$ are almost identical for the ghost
conditions IC1 and IC2, and they are much larger than for IC0.  For the 
non-ghost condition IC0 the energies steadily fall until very late times.
For each of the two ghost conditions the energy increases to the point 
(about $z = 4.7$) at which the cosmological redshift begins to dissipate 
it. It must be remembered that these results follow from the {\it linearized}
field equations

\begin{figure*}[htbp]
\begin{center}
 \begin{tabular}{cc} 
  \includegraphics[width=0.45\textwidth]{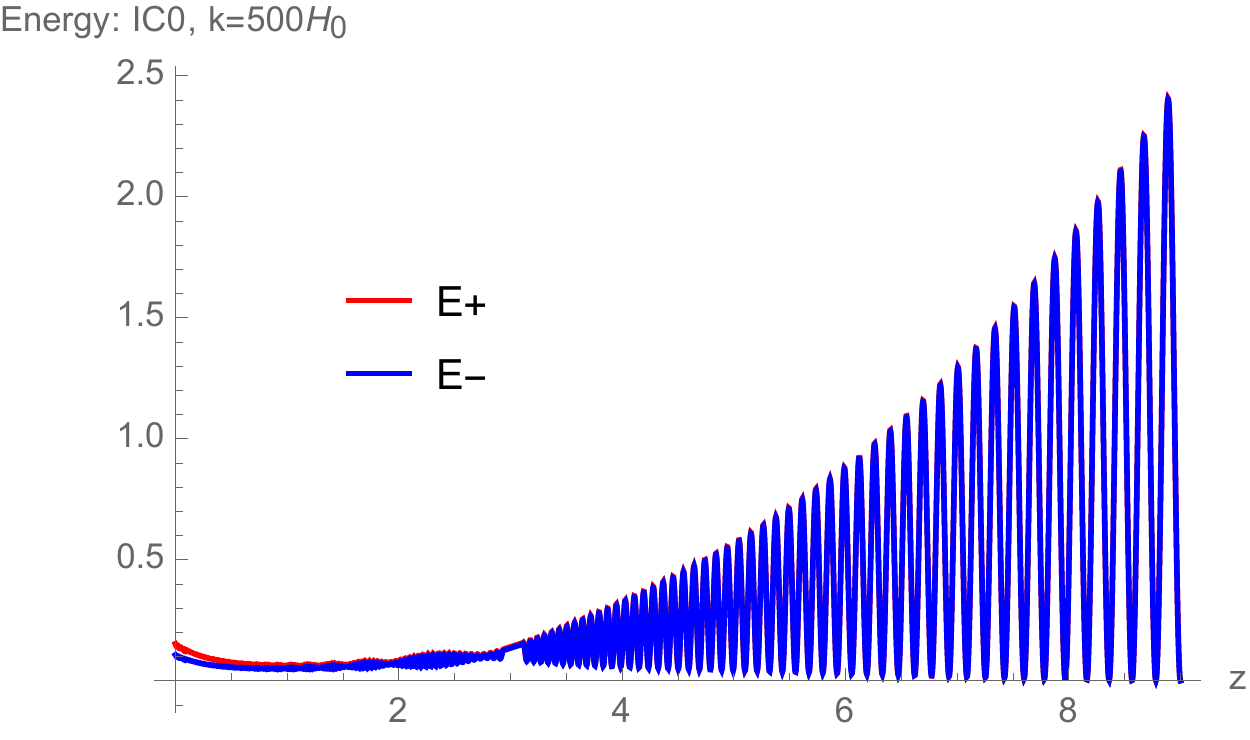}
  &
  \includegraphics[width=0.45\textwidth]{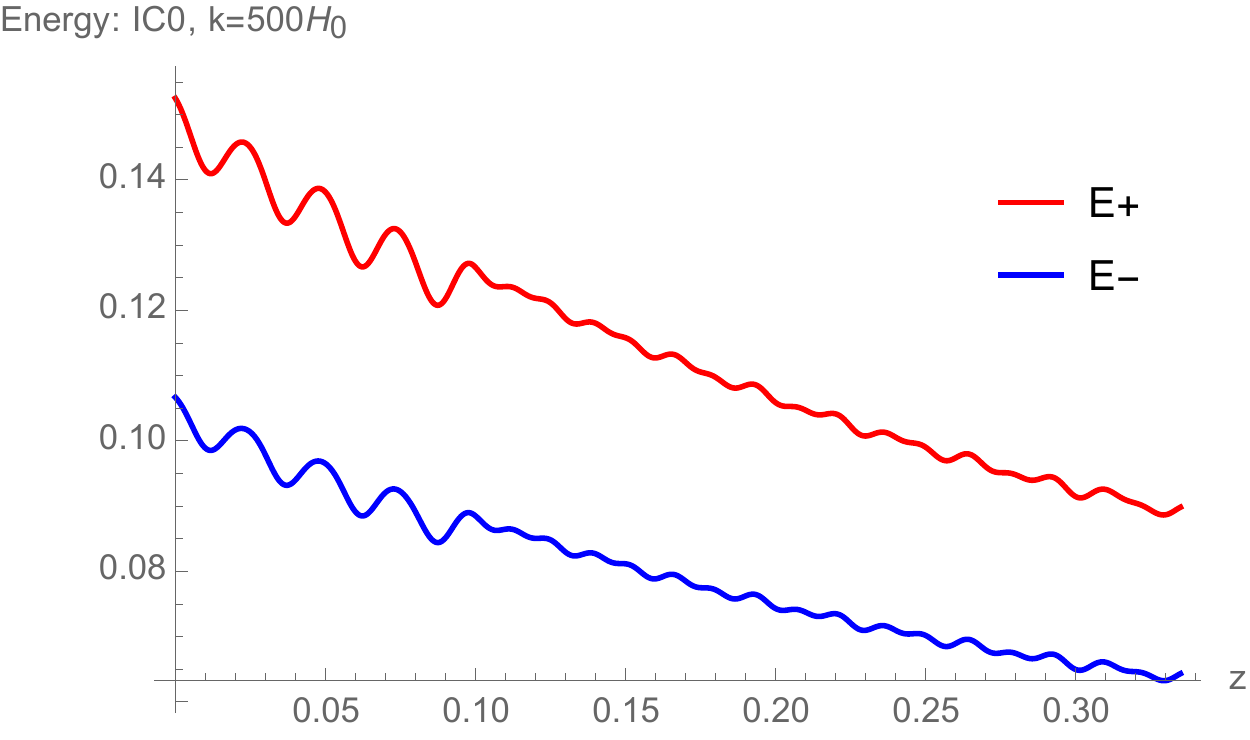}
\end{tabular}
\end{center}
\begin{center}
 \begin{tabular}{cc} 
  \includegraphics[width=0.45\textwidth]{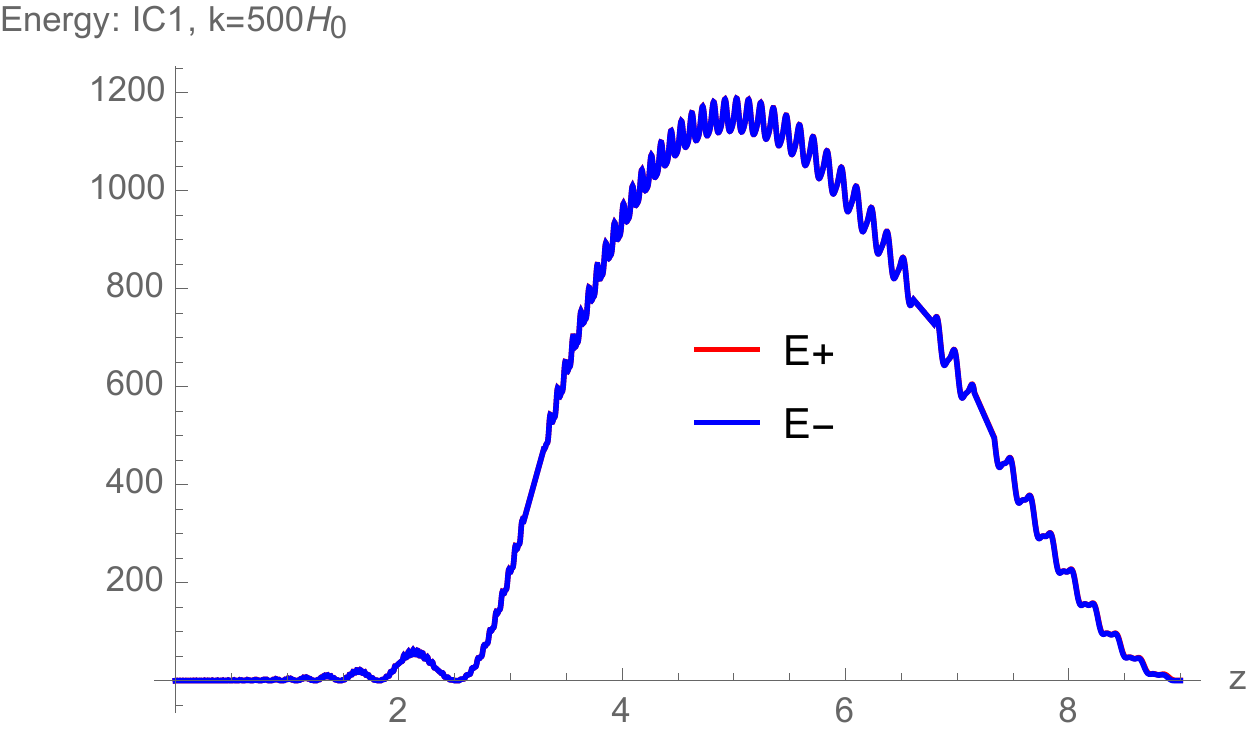} 
  &
  \includegraphics[width=0.45\textwidth]{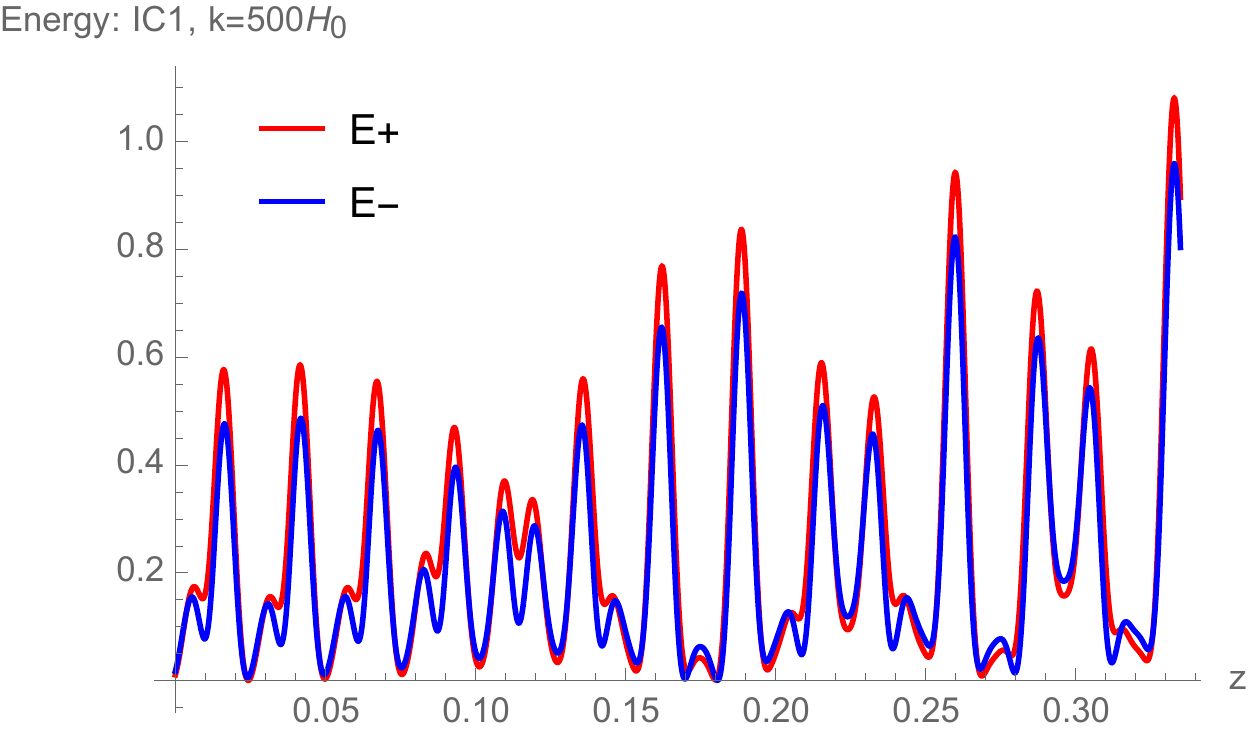} 
 \end{tabular}
\end{center}
\begin{center}
 \begin{tabular}{cc} 
  \includegraphics[width=0.45\textwidth]{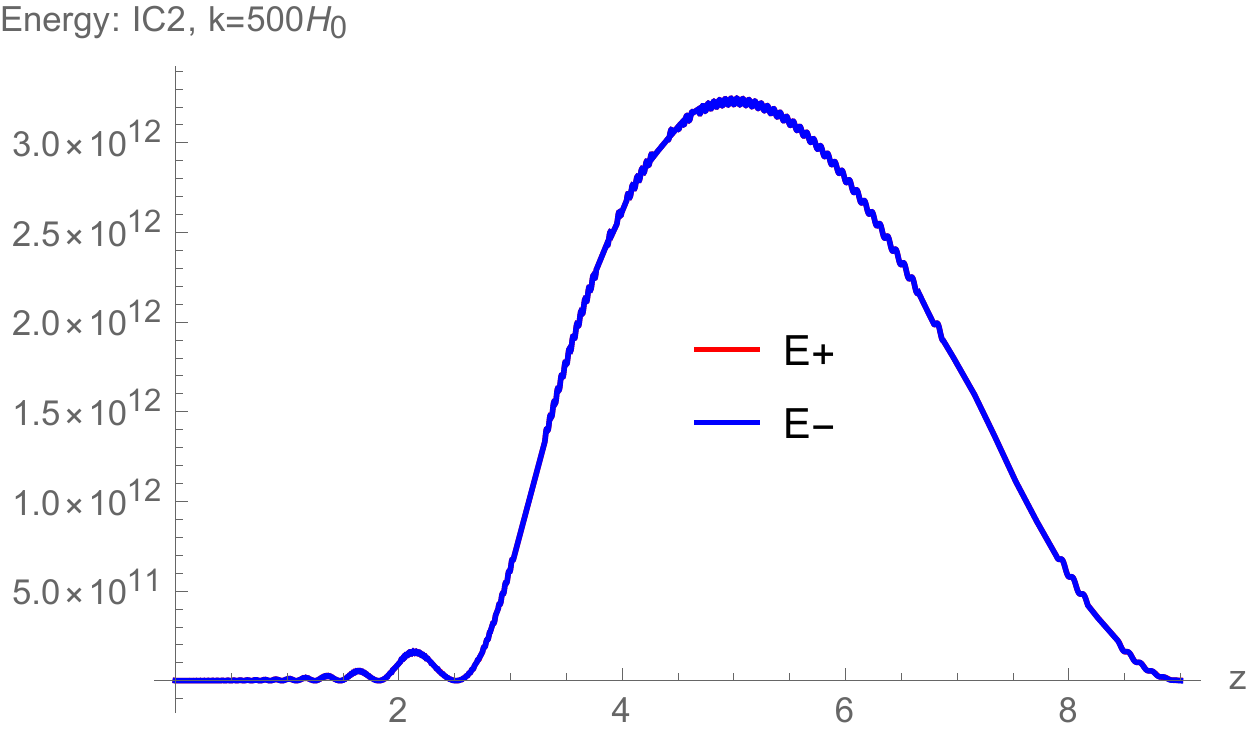}
  &
  \includegraphics[width=0.45\textwidth]{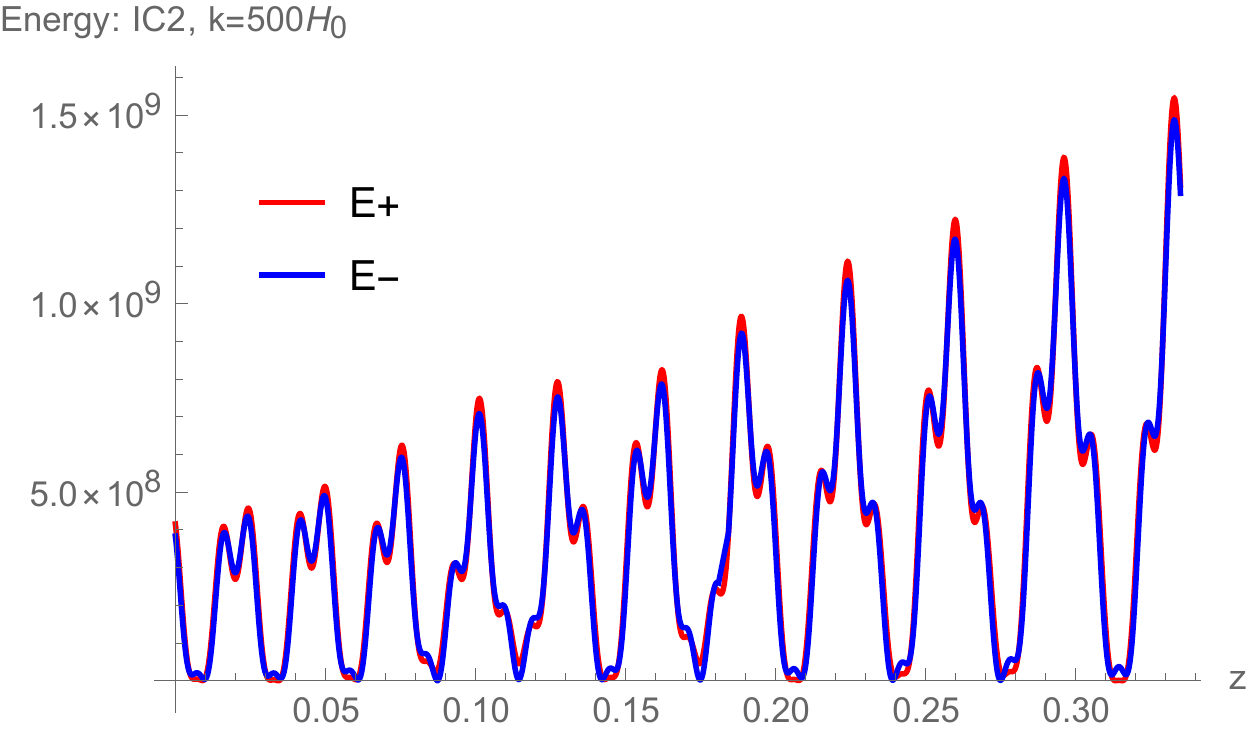}
 \end{tabular}
\end{center}
\caption{The $k = 500 H_0$ kinetic energies $E_{\pm}$ (\ref{Epm}) versus 
redshift $z$ for the initial conditions IC0 (\ref{nonlocal-IC}), IC1 
(\ref{IC1}), and IC2 (\ref{IC2}). In each case the left hand graphs show 
the full range $0 < z < 9$, whereas the right hand graphs provide an 
expanded view of the late time regime $0 < z < 0.34$.}
\label{fig:k500-energy-z}
\end{figure*}

The peak in $E_{\pm}$ comes earlier, and is much higher, for larger wave 
numbers $k$. Figure~\ref{fig:k500-energy-z} shows the result for $k = 500 H_0$, 
at which the peak occurs at about $z = 5$. We confirmed the general trend
by runs at $k = 300 H_0$, $k = 700 H_0$ and $k = 1000 H_0$, but there is
no point in presenting these graphs.

\section{Discussion}

Nonlocal cosmology (\ref{DW-action}) is not an attempt to replace general
relativity but rather to provide a phenomenological representation of quantum
infrared effects which grew nonperturbatively strong during the epoch of
primordial inflation. The idea is that general relativity is the fundamental
theory of gravity, but what we observe is the nonlocal effective field equations, 
just as quantum electrodynamics is the fundamental theory of charged matter, but
the observed running of charge manifests in solutions to the nonlocal effective 
field equations. The degrees of freedom of nonlocal cosmology are the same as 
those of general relativity \cite{DW-2013,W-review-2014}. This is apparent from 
the fact that the inverse scalar d'Alembertian is defined with retarded boundary
conditions on an initial value surface corresponding roughly to the epoch of
primordial inflation. 

In sharp contrast, the localized model (\ref{DW-action-localized-IBP}) 
represents an alternate gravity theory in which two fundamental scalars 
figure. One of these scalars is a ghost, which means the localized theory 
suffers from a virulent kinetic instability that causes the ghost to be 
more and more highly excited, with a consequent excitation of the positive 
energy degrees of freedom. That is apparent from the relative signs of the
scalar stress tensors (\ref{stress}). Hence the localized model cannot 
possibly be acceptable. The worrisome thing for nonlocal cosmology is that 
one can view its Lagrangian (\ref{DW-action}) as a constrained version of 
(\ref{DW-action-localized-IBP}) in which the two scalars and their first 
time derivatives vanish on the initial value surface. (Note that this
constraint already precludes the worst instability of having arbitrarily
large values of $\delta A_{+}(0,\vec{k}) = \delta A_{-}(0,\vec{k})$ at
arbitrarily high wave numbers.) That does not necessarily condemn 
nonlocal cosmology to suffer the kinetic instability; the familiar 
conformal factor of unmodified general relativity would also be a 
ghost were it not for the Hamiltonian constraint. But it is prudent to 
check that the constraint of nonlocal cosmology is effective in 
controlling the ghost.

Note that the constraint does not compel the ghost field to remain
zero, any more than the constraint of general relativity requires
the conformal factor to remain unity. Both the ghost mode of nonlocal
cosmology and the conformal factor of general relativity evolve even
in the cosmological background. What we seek to show is rather that
the constraint protects against explosive growth.

If we had an energy functional for nonlocal cosmology the check 
would be simple. In the absence of such an energy functional we have 
instead studied the evolution of linearized spatial plane wave 
perturbations about the cosmological background, both starting from 
the retarded boundary conditions (\ref{nonlocal-IC}) of nonlocal
cosmology and with more general initial conditions (\ref{IC1}) and 
(\ref{IC2}). In Fig.~\ref{fig:ic0-solutions} we see that the 
perturbations of nonlocal cosmology show no sign of the kinetic 
instability. Although the scalar $\delta X(t,\vec{k})$ does experience 
some decaying oscillations at early times, they are not communicated 
to the other fields. Evolutions from more general conditions are shown 
in Figures \ref{fig:ic1-solutions} and \ref{fig:ic2-solutions}. In both 
cases the oscillations of $\delta X(t,\vec{k})$ are much larger, they 
grow, and they are communicated to the other perturbation fields. This 
is how a kinetic instability manifests.

Figure~\ref{fig:amplitude-z-t} gives the ghost and normal scalars, $\delta A_+$ and $\delta A_-$,
respectively, for the three initial conditions. With retarded boundary conditions
(IC0) the two experience some decaying oscillations at first and go on to distinct 
evolutions at late times. For the other boundary conditions (IC1 and IC2) the
oscillations are much larger, they grow, and they are coupled. Recall that the ghost
dragging along the other fields is what characterizes a kinetic instability. 
Figures \ref{fig:energy-z} and \ref{fig:energy-time} show the same thing using the 
magnitudes of the kinetic energies.

One thing we cannot do with the linearized field equations is exhibit the explosive
instability associated with mixing from different wave vectors, when each one starts
with general initial value data. However, within the limitations of what is easy to
study numerically, our analysis has provided strong evidence against nonlocal 
cosmology suffering from the kinetic instability of its localized cousin. It also 
demonstrates why the localized version is so problematic.

Devising a full stability proof would require an energy functional, which does not 
exist for gravitating systems in cosmology. We suspect that this may not be as big 
an obstacle as it might seem because nonlocal cosmology approaches de Sitter at late 
times. So we propose adapting the famous result of Abbott and Deser \cite{Abbott} for 
general relativity with a positive cosmological constant. Instead of the Hilbert 
action we would use the localized Lagrangian (\ref{DW-action-localized-IBP}) with the 
scalars constrained to obey retarded initial conditions. And instead of de Sitter
providing the asymptotic conditions, it would be the background solution for nonlocal 
cosmology. Then we would try to prove positivity of the energy for sub-horizon 
fluctuations, just as Abbott and Deser did. That seems a worthy project for the future.

\section*{Acknowledgements}

We are grateful for correspondence on this subject with S. Deser. 
This work was partially supported by the European Research Council 
under the European Union's Seventh Framework Programme 
(FP7/2007-2013)/ERC Grant No. 617656, ``Theories and Models of the 
Dark Sector: Dark Matter, Dark Energy and Gravity"; by NSF grants 
PHY-1506513 and PHY-1806218, and by the Institute for Fundamental 
Theory at the University of Florida.

\end{document}